\documentclass[12pt]{article}
\date{ \today }
\usepackage[dvips]{epsfig}
\begin{document}
\begin{center}

{\bf
{\Large Thermodynamical features of multifragmentation in peripheral
 Au + Au Collisions at 35 A.MeV}
}

\vspace{0.5cm}

\today

\vspace{0.5cm}
\vspace{0.5cm}

\noindent
{M.~D'Agos\-ti\-no, A.~S.~Botvina
\footnote{\scriptsize Institute for Nuclear Research, Russian Academy of 
Science, 117312 Moscow, Russia}
, M.~Bruno}

{\scriptsize Dipartimento di Fisica and INFN, Bologna, Italy}

\vspace{0.5cm}

{A.~Bonasera}

{\scriptsize INFN, Laboratorio Nazionale del Sud, Catania, Italy}

\vspace{0.5cm}

{J.~P.~Bondorf, I.~N.~Mishustin
\footnote{\scriptsize Kurchatov Institute, Russian Scientific Center, 
123182 Moscow, Russia}}

{\scriptsize Niels Bohr Institute, DK-2100 Copenhagen, Denmark}

\vspace{0.5cm}

{F.~Gulminelli, R.~Bougault, N.~Le~Neindre}

{\scriptsize LPC, IN2P3-CNRS, ISMRA and Universit\'e, F-14050 Caen Cedex, 
France}

\vspace{0.5cm}

{P.~D\'esesquelles}

{\scriptsize GANIL, CEA, IN2P3-CNRS, B.P. 5027, F-14021 Caen Cedex,
and Universit\'e Joseph Fourier, Grenoble, France}

\vspace{0.5cm}

{E.~Geraci, A.~Pagano}

{\scriptsize INFN Sezione di Catania, Catania, Italy}

\vspace{0.5cm}

{I.~Iori, A.~Moroni}

{\scriptsize Dipartimento di Fisica and INFN, Milano, Italy}

\vspace{0.5cm}

{G.~V.~Margagliotti, G.~Vannini}

{\scriptsize Dipartimento di Fisica and INFN, Trieste, Italy}

\vspace{0.5cm}

\end{center}

\newpage

\small
\begin{abstract}
\baselineskip=14pt
The distribution of fragments produced in events involving the
multifragmentation of excited sources is studied for peripheral
Au + Au reactions at 35 A.MeV. 

The Quasi-Projectile has been reconstructed from its de-excitation products.
An isotropic emission in its rest frame has been observed, indicating that an
equilibrated system has been formed. The excitation energy of the 
Quasi-Projectile has been determined via calorimetry.

A new event by event effective thermometer is proposed based on the energy 
balance.
A peak in the energy fluctuations is observed related to the heat capacity 
suggesting that the system undergoes 
a liquid-gas type phase transition at an excitation energy $\sim 5 \ A.MeV$
and a temperature $4 - 6 \ MeV$, dependent on the freeze-out hypothesis.
By analyzing different regions of the Campi-plot, the events associated 
with the liquid and gas phases as well as the critical region are
thermodynamically characterized. 

The critical exponents, $\tau, \beta, \gamma$, extracted from the high moments
of the charge distribution are consistent with a liquid-gas type phase 
transition.

\vspace{0.5cm}
\end{abstract}

\newpage

\baselineskip=20pt

\section{Introduction}
Nuclear multifragmentation and its possible connection to the occurrence of a
phase transition of the liquid-gas type has been the subject of intensive
theoretical and experimental 
investigations~\cite{curt,jaq,purd,rapis,Gulm_phil,Bib_Bon2,grossreport90}.
Theoretical studies indicate that infinite nuclear matter has an equation of
state similar to that of a Van der Waals gas which is characterized by the
existence of a liquid-gas phase transition. Moreover, recent experimental
results show strong evidence for this occurrence in fragmenting nuclear 
systems. In particular the {\it ALADIN} collaboration has 
measured a caloric 
curve resulting from the fragmentation of 
the Projectile and Target Spectators formed in peripheral collision Au + Au at 
600 and 1000 A.MeV, and found a behaviour expected for a first order 
liquid-gas phase transition~\cite{aladin,aladincentral}.
The {\it EOS} collaboration has also measured a caloric curve and extracted 
the critical exponents of fragmenting nuclear systems produced in the 
collision of 1 A.GeV Au nuclei with a Carbon target, finding values 
consistent with the expected values at the critical point~\cite{eos}. 

To experimentally study a nuclear system undergoing a phase transition
events coming from the decay of a single source in a wide 
range of excitation energies have to be considered, so that it is possible to 
explore different kinds of de-excitation, from evaporation to 
multifragmentation towards vaporization.
One can then hope to cross the critical region~\cite{Gulm_phil,campilattice}
where the system manifests signals of a phase transition.

In addition, the size of the studied source should be as large as possible
to minimize finite size effects; and the collective radial flow (due to the 
compression of the nuclear matter in the first stages of the reaction), as well 
as the rotational energy of the emitting system should be as small as 
possible.

In central collisions, with reactions involving heavy projectiles and targets,
it seems possible to overcome the difficulty concerning the size of the 
system. Indeed in Au + Au collisions at 35 A.MeV the {\it Multics - Miniball}
collaboration has 
studied~\cite{Bib_DAg3,Bib_DAg4,Bib_DAg1,pierre} the characteristics of an 
equilibrated {\it composite} system with a size larger than 300 nucleons.
Also in the Xe + Sn reactions the {\it Indra} 
collaboration~\cite{indra,Bib_Bou2} examined 
an intermediate system with a size similar to the Au one.
In these collisions, however, when the system is well identified, its excitation
energy is distributed in a narrow range around the mean value so that 
it is impossible to build an excitation function with these data, except
if the same collision is studied varying the beam energy 
in an almost continuous way. Even so, the collective energy contribution 
to the excitation energy depends strongly on the incident energy and it is not 
a trivial process to disentangle this component.

Peripheral collisions seem to give a better chance to study the critical 
region, when selecting 
the Quasi-Projectile (QP) and Quasi-Target (QT) in the final state. 
In the reaction Au + Au  at 35 A.MeV the maximum available centre of mass 
energy per nucleon is $8.71 \ A.MeV$ which determines the scale of the QP and 
QT excitation energies. In principle one can observe 
sources with an excitation energy from a few hundred of A.keV to several units 
of A.MeV. Moreover, the radial collective energy is limited to 
$\sim 1 \ A.MeV$ for the most central collisions~\cite{Bib_DAg3}, and 
should be smaller in peripheral and semi-peripheral collisions.
The contribution of rotational energy can be evaluated in the multifragmentation
region by comparisons with thermal predictions.

Events associated to the Quasi-Projectile decay were already experimentally 
studied by looking for signals of a critical behaviour~\cite{letter,subs} and 
anomalies in the caloric curve~\cite{prctrieste}.
Several indications in favour of the occurrence of a critical behaviour were 
found by the study of the moments of the charge 
distributions~\cite{letter,subs}. They have been revealed mainly by the 
particular shape of the Campi scatter plot and by the presence of large 
fluctuations. 
Similar analyses were also performed~\cite{auger} for peripheral collisions 
on the $Xe + Sn $ $50 \ A.MeV$ reaction. 
In the caloric curve analysis of Ref.~\cite{prctrieste}, temperatures $T_{iso}$ 
(calculated from several double isotope ratios) increasing from 3.7 to 4.5 MeV 
with decreasing impact parameters were observed.
These values resulted in agreement with Aladin and EOS data in the common 
range of excitation energy.

Since the investigation of phase transitions implies the study of 
observables related to the critical behaviour on an event by event 
analysis~\cite{cmdtheo,theomoh,gross1,campi}, 
the temperatures, as well as the excitation energies measured so far, cannot 
be used as the critical parameter as they are known only on the average.
Indeed the experimental determination of these observables has been 
performed up to now in bins of other measured observables, such as 
$Z_{bound}$~\cite{aladin,aladincentral},
charged particle multiplicity~\cite{prctrieste,huang,xi,xib,hauger} and 
transverse energy~\cite{ma}.

In addition, the value of the measured temperatures could result from an 
average of different stages of the de-excitation~\cite{Bib_Bou2,xib}.
In the framework of the statistical model SMM~\cite{Bib_Bon2}, for instance, 
it has been shown that decaying systems with freeze-out temperatures of about 
$6 \ MeV$ give final isotopes with apparent double ratio temperatures 
on the order of $4 \ MeV$, reasonably approaching the experimental 
values~\cite{prctrieste}.
Only when considering excitation energies much higher~\cite{bond98} than those 
here considered the isotopic ratios give a good approximation of the
freeze-out temperature.
Other ambiguities arise from very recent experimental results on the 
temperature determination through the excited state 
populations~\cite{li5aladin}.
No effort is here made to investigate the reasons of the
discrepancies between results obtained with different
isotopic and excited states thermometers and
to establish which are the most reliable ones. (For a discussion on this
topic, see for instance Ref.s~\cite{aladincentral,Gulm,siwek}).

In this paper we investigate in detail the disassembly of the QP
formed in peripheral $^{197}$Au + $^{197}$Au 35 A.MeV collisions.
The data selection method guarantees in each event
a unique emitting source of the fragments.

A new effective event by event thermometer is proposed which allows for 
the first time the extraction of a "primary" caloric curve by backtracing the 
experimental information to the freeze-out configuration. It also allows for
the computation of new thermodynamical observables, such as an effective heat 
capacity, through the energy fluctuations.
A detailed analysis of the moments of the charge distribution is 
performed on an event by event basis.

\section{Experiment}
The experiment was performed at the National Superconducting Cyclotron
Laboratory of the Michigan State University.
Beams of Au ions at $E/A = 35 \ A.MeV$ incident energy from
the K1200 cyclotron were used to bombard Au foils of approximately
$2.5 \frac{mg} {cm^2}$ areal density.

Light charged particles and fragments with charge up to the beam charge were 
detected at $\theta_{lab}$ from $3^{\circ}$ to $23^{\circ}$ by the 
{\it Multics} array~\cite{mcs}, with an energy threshold of about 1.5 A.MeV, 
nearly independent on fragment charge.
Light charged particles, $Z=1$ and $Z=2$ isotopes and fragments with charge up 
to $Z=20$ were fully identified by 160 phoswich detector elements of the MSU 
{\it Miniball} \cite{mini}, covering the angular range from 
$23^{\circ}$ to $160^{\circ}$.
The charge identification thresholds were about $2, \ 3, \ 4 \ A.MeV$
for $Z = 3, \ 10, \ 18$, respectively.

The geometric acceptance of the combined array was greater than 87\% of 
$4\pi$. Dynamical calculations indicated~\cite{theomoh}  that the apparatus
is able to detect all the generated events up to a maximum impact parameter 
$b \approx 11 \ fm$, with the experimental triggering condition (at least
two fired detectors). Higher impact parameters $b= 12 - 14 \ fm$ are 
detected with lower efficiency, being the Quasi-Target products 
below the energy thresholds and the Quasi-Projectile ones mainly flying at
laboratory angles smaller than the minimum detection angle.

We have studied in detail the efficiency for QP detection 
by performing calculations with the statistical SMM
model~\cite{prctrieste} for an Au source, excitation energies ranging up 
to 9 A.MeV and freeze-out density $\rho_0/3$ ($\rho_0$ stands for the 
normal nuclear density).
For the velocity of the source in the laboratory frame, the value $v_{QP}$ 
was assumed from energy conservation:
\begin{equation}
v_{QP} = \sqrt { 2 \ (\epsilon_{c.m.} - \epsilon^*_{QP}) } + v_{c.m.}
\label{conservation}
\end{equation}
where $\epsilon_{c.m.}$ is the available centre of mass energy per nucleon,
$v_{c.m.}$ the velocity of the centre of mass of the reaction in the laboratory 
reference frame and $\epsilon^*_{QP}$ is the QP excitation energy.

The predictions were filtered through the software
replica of the apparatus, taking into account the geometry, energy 
thresholds, Z-identification limitations and multiple hits.
It resulted that for events where the total detected charge 
is at least 70\% of the projectile charge, the event intrinsic efficiency
for the detection of the QP products
(ratio between the charged particle multiplicity detected by the
apparatus and the one before the filter)
is on the average $70\%$ and that 
the ratio between the total charge bound in form of fragments after and
before the filter
is greater on the average than 93\% (96\%) for fragments with charge 
$Z \ge 3 \ (6)$,
respectively, over the whole range of the excitation energy $\epsilon^*$.

For these reasons the analysis presented in this paper is essentially based
on the detected fragments. We will show in the next Section that, in our range
of excitation energy, the limited efficiency for light-particle detection 
does not affect the calculation of observables which characterize the
decaying systems.

\section{Data Selection and QP characteristics}
As in the previous analyses of peripheral collisions~\cite{letter,subs} we 
selected the most peripheral collisions by requiring the velocity of the 
largest fragment in each event to be at least 75\% of the beam velocity.
After a shape analysis, i.e. an analysis of the variance of the fragment 
velocities~\cite{Bib_DAg4,cugnon},
the other IMFs of each event were considered as belonging to the QP if forward
emitted in the ellipsoid reference frame.
Only events where at least 70\% of the Au charge was detected in the
forward hemisphere of the ellipsoid were accepted for the QP source 
reconstruction. The velocity of the source was then calculated as the weighted 
mean of the IMF velocities. The last step is to add the light particles 
to the source. This has been done by doubling the contribution of 
light particles forward emitted in the source reference frame to 
minimize the contribution of pre-equilibrium particles~\cite{peter,cussol}.

In Fig.~1~a),~b) and~c), we display the probability spectrum of 
$cos(\theta_{flow})$, where $\theta_{flow}$ is the angle formed by the main 
eigenvector resulting from the shape analysis and the beam axis.
Different panels correspond to different selections applied to the data.
Fig.~1~a) corresponds to all the collected events,
Fig.1~b) to events with the constraint on the velocity of the
largest fragment and finally Fig.1~c) represents the $cos(\theta_{flow})$ 
spectrum for the events satisfying both the conditions on
the velocity of the largest fragment and on the total charge detected in
the forward hemisphere. 

In Fig.~1~d),~e) and ~f) a recently proposed 
observable~\cite{lecolley,galichet} is presented, which allows for the 
visualization of the source(s) in the analyzed events.
The ensemble averaged {\it charge density}
$\langle \rho_Z (v_{par}) \rangle $ is defined as:
\begin{equation}
\langle \rho_Z (v_{par}) \rangle = \langle 
\frac {\Sigma Z(v_{par})} {\Sigma Z} \rangle 
\label{chargedensity}
\end{equation}
where $\rho_Z (v_{par})$ represents the event by event distribution in the 
velocity $v_{par}$ of the collected charge fraction.
In Ref.~\cite{galichet} the preferred axis to project fragment velocity vectors
was the main eigenvector of the ellipsoid reference frame, in our case the axis 
chosen is parallel to the QP velocity. 
The sums in the numerator and denominator of equation~(\ref{chargedensity})
have been limited to fragments ($Z \ge 3$). This is because light particles,
symmetrized with respect to the QP velocity, do not contribute to the source
identification.
For our events this observable thus represents the distribution of the 
collected charge bound in fragments along the direction of the QP velocity.

Panels d), e) and f) of Fig.~1 correspond to the same selections performed
for the plots shown in Fig.~1~a), ~b) and ~c), respectively.
In all these panels a clear contribution from QP emission
is seen, while the Quasi-Target products (below the energy thresholds)
are lost, even if fast moving forward emitted light fragments can contribute 
to the observed distribution.
The fragment emission in the mid-rapidity region decreases while applying the 
constraints on the largest fragment velocity and on the total charge detected in
the forward hemisphere. By analyzing through gaussian fits~\cite{Indrabormio98}
the $\rho_Z$ distribution of Fig.~1 ~f) in different bins of impact parameter, 
we found indication that the possible pollution of fragments not 
belonging to the QP goes from about 2\% for the most peripheral collisions
to less than 10\% for the most central ones.

The distribution of the charge of the reconstructed QP ($Z_{QP}$) is shown in 
Fig.~2~a). 
For the analysis presented in the following Sections we selected events
where $Z_{QP}$ is within $\pm$ 10\% of the Au charge (shaded area). 
For these events the mean measured total parallel momentum is 79\%
of the projectile linear momentum with a standard deviation 7\%.
Fig.~2~b) represents the yield of different reduced impact parameters 
$b/b_{max}$ after the selections leading to the plots of Fig.~1~c) and ~f) 
and the shaded area of Fig.~2~a).
The ratio $b/b_{max}$ has been calculated as in Ref.~\cite{tsang} through the 
detected charged particle multiplicity before the selection process except for
a minimum bias trigger (at least two fired detectors).

Being that the analyzed events are mainly peripheral with small contaminations 
from the mid-rapidity source and {\it well detected} from the 
point of view of the size of the $QP$ and the linear momentum, 
the excitation energy has been calculated via calorimetry.
Possible contributions of non thermal origin to the excitation energy have been
evaluated by comparing the measured kinetic energies to thermal predictions. 
Indeed some pre-equilibrium light particles or fragments could have been 
included in the reconstructed Quasi-Projectile and contributed to its 
excitation energy. 

The excitation energy $E^*$ of the Quasi-Projectile in each event results 
from the energy balance between the initial stage at the freeze-out and the 
final stage of detected fragments:
\begin{equation}
m_0 + E^* = \left[ \sum (m_j + E_j ) + \sum (m_n + E_n ) \right].
\label{calorimetry}
\end{equation}
Here $m_0$ is the mass of the QP, determined from the reconstructed charge
assuming a charge-to-mass ratio as in the entrance channel.
$m_j$, $E_j$ $(j=1,\dots,N_c)$ are the masses and the kinetic energies (in the 
source reference frame) of the charged products (multiplicity $N_c$),
$m_n$, $E_n$ are the masses and the kinetic energies 
of neutrons.
The $m_j$ were calculated from the measured $Z_j$ through the 
numerical inversion of the Epax parameterization~\cite{epax}, except for 
charges $Z=1, \ 2$ where the measured values were used.
$m_0,\ m_j,\ m_n$ take into account the mass excess.

The number of free neutrons was obtained as the difference
between the number of nucleons ($A_0$) of the QP and the sum of nucleons 
bound on the detected particles and fragments ($N_n = A_0 - \sum A_j$).
The average kinetic energy of neutrons was estimated with different 
approximations which lead to values of the excitation energy varying 
less than 5\%.
These estimates were done:
\begin{description}
\item {i)}
by assuming for $E_n$~\cite{hauger} a Maxwell-Boltzmann thermal distribution, 
consistent with volume emission~\cite{wada}.
In equation~(\ref{calorimetry}) 
$\sum E_n \ = \ \frac{3}{2} N_n T = \ \frac{3}{2} N_n \sqrt{E^*/a}$
has been assumed, where 
{\it a} = A/13~\cite{hagel} represents the level density parameter of the 
degenerate Fermi gas. 
\item{ii)}
by assuming that in each event neutrons have 
the same mean kinetic energy as protons but decreased by the Coulomb barrier
from the $QP$ source. 
\item{iii)}
by assuming, similarly to SMM model, that the Coulomb barrier depends on the 
thermal excitation energy of the emitting system (the light particle
emission switches from surface evaporation to volume emission with increasing 
excitation energy).
This requires applying a two-step calculation to the data.
First we calculated the proton barrier with the estimate of $E^*$ described in
{\it i)} and then, by solving equation~(\ref{calorimetry}),
we obtained the final value of $E^*$.
\end{description}

As already mentioned, possible contributions of non thermal origin 
were evaluated by a model comparison (see next Section).
They resulted smaller than 10\% in the 
whole range of the $QP$ excitation energy.
In the following, unless differently stated, we will use only the thermal 
part of the excitation energy i.e. the value obtained via calorimetry, reduced 
by the nonthermal contribution.
We will show however, that in our excitation energy range, possible distortions 
introduced by the non thermal component do not change our main results and 
conclusions.

Equation~(\ref{calorimetry}) was checked by applying it to SMM events
and by comparing the obtained values to the input of the model. 
We found that calculated values differ from the input by no more than 
$3\%$ both when applying 
eq.~(\ref{calorimetry}) to filtered or not filtered events.
From this result we can conclude that, in our energy range, a high efficiency 
for fragment detection is essential for a correct determination of the 
excitation energy, while some inefficiency for light particle detection only 
slightly affects the calculation.

In Fig.~2~c) we report the $\epsilon^* = E^*/A_{QP}$ experimental distribution. 
The weight of the impact parameter on the experimental yield is evident. 
Therefore we will perform the study of the source characteristics 
and other observables in bins of $\epsilon^*$ 
(or $b/b_{max}$) narrow enough to safely neglect the weight of the impact
parameter within each bin.

In Fig.~2~d) the experimental correlation between the laboratory velocity of 
the QP and its excitation energy is shown. The line represents the values 
obtained from equation~(\ref{conservation}). The arrows represent the values
of the beam velocity and of the centre of mass velocity in the laboratory 
reference frame. The experimental values are in agreement with the
expected values, apart from some deviation at high excitation energy.
Indeed our event selection could have partially eliminated events where the 
Quasi-Projectile was slowly moving with respect to the other fragment sources.

To deeper investigate whether the QP can be considered as the unique source of
the observed fragments, we have studied the charge density
$\langle \rho_Z \rangle$ (equation~(\ref{chargedensity})) in different bins
of excitation energy (Fig.3).
We should to remark that recent theoretical studies~\cite{botv_gross} have
shown that some deformation in the velocity spectra of light fragment can 
be ascribed to the Coulomb influence of the second source (QT) on the 
decay of the QP. This prediction could also explain the slight deformation of
$\langle \rho_Z \rangle$ observed in Fig.3~d).
For the first three panels deviations of $\langle \rho_Z \rangle$ from a 
symmetric shape are negligible, since at low excitation energy the multiplicity 
of light fragments is low. At higher excitation energies (panel ~d))
the tail of the distribution is slightly stretched in the backward 
direction, as expected when light fragments minimize 
the potential energy through their position in the freeze-out volume.

In conclusion the distributions shown in Fig.~1 and the QP characteristics 
shown in Fig.~2 and 3 make us confident that the utilized constraints 
do, in, select the events where the QP break-up 
is the dominant mechanism of fragment production.

\section{Comparison with a statistical model}
To verify the extent to which the reconstructed QP can be considered 
as the unique fragment source for the selected events, we compared 
the experimental Z-distributions with the predictions of a statistical model.

The detailed description of the Statistical Multifragmentation Model (SMM) 
can be found in Ref.\cite{Bib_Bon2}. Here we only emphasize that the 
basic assumptions of the model are the equilibration of the decaying 
system and the statistical distribution of the probabilities of the 
break-up channels. 

In Ref.~\cite{prctrieste} it was already shown that SMM can describe 
rather well the mean elemental event multiplicity $N(Z)$ for 
$Au$ $+$ $Au$ $35 \ A.MeV$ peripheral collisions.
Here we want to look in more detail at the degree of reproduction of
other observables.
In principle, the detailed reproduction of the 
variances of the observables distributions should require the use of 
sophisticated backtracing procedures~\cite{pierre} which work in the 
multivariate space of the source characteristics. But this extraction of 
the sources distribution is beyond the aim of this paper.
So we simplify our analysis by comparing experimental
observables with SMM predictions coming from fixed values of the source size
($A_0 = 197, Z_0 = 79$, density $\rho_0/3$) and for excitation energies 
continuously varied up to 9 A.MeV (8.71 A.MeV is the c.m. available energy 
per nucleon for our reaction).
For the velocity of the source in the laboratory frame the values from energy 
conservation were assumed (eq.~\ref{conservation}) .

In Fig.~4 we present the mean elemental event multiplicity distribution
$N(Z)$ for several bins of impact parameter.
No relative normalization on the presented histograms was performed.
The distributions have been only normalized to the total number of events
both for data and SMM predictions.
The solid line in Fig.~4 represents filtered SMM predictions, while the
dashed one corresponds to not filtered ones. One can see that no distortions 
are introduced by the filter, since the apparatus has a quite high efficiency
for QP products (well above the energy thresholds).
With the considered QP ensemble the model reproduces quite
well the experimental distributions, apart from the region of Fission 
Fragments at the most peripheral impact parameters.
In particular it seems that the calculation gives less fissions 
and a more asymmetric mode with respect the data.  
However this is a particular channel existing at low excitation energy 
(about 2 A.MeV for the experimental data).
Moreover the charge asymmetry between the two fission fragments can be 
influenced by a rotational motion of the emitting system~\cite{largemom} not
considered in the present calculation. 

By removing both from data and model predictions the Fission Fragments (Fig.~5)
it is clearly seen that the agreement of $N(Z)$ is improved and that 
also the event charge partition, i.e. the distribution of $N(Z)$ for the 
heaviest fragments in each event, is very well reproduced.
In Fig.5 we show experimental events corresponding to the impact parameter range
$0.8 < b/b_{max} \le 0.9$ but similar results have been obtained for all the 
$b/b_{max}$ intervals shown in Fig.~4.

So we believe that this limited lack of agreement between data and SMM is
well understood and does not affect the {\it multifragmentation} region
which is of interest for the analyses of this paper.

As mentioned in the previous Section SMM predictions were also used, in each 
bin of impact parameter and for each charge, to evaluate
possible contributions coming from pre-equilibrium light particles and
fragments and/or from some collective energy.
This was done by comparing the measured and predicted kinetic energies per 
nucleon. From Fig.~6 it is evident that, while for heavy fragments the
agreement between data and predictions is excellent, for fragments
with charge smaller than 10 an extra kinetic energy is present in
the data relative to the thermal+Coulomb contribution. 
These differences have been taken into account to evaluate the non
thermal contributions to the calorimetric excitation energy. 

In Fig.~7 the experimental mean calorimetric excitation energy is compared
with the SMM mean values ($\pm$ the standard deviation) needed to reproduce 
the experimental charge distribution and the charge partition.
For this comparison we eliminated fission events both from data and 
predictions.
For all the bins of the impact parameter a very good agreement is evident
either including or excluding the collective energy.

Our experimental partitions show a high degree of thermalization, apart from a 
small contribution of extra-kinetic energy for light charged particles and small
fragments. Successive applications of thermalization hypothesis to 
the data should take into account that some nucleons and light particles can
have escaped the system through surface emission before the thermalization.
Even if this effect should be more important at excitation energies higher than
those here considered in our following analysis it will be taken into account.

The high degree of reproduction of the observables in each 
bin of impact parameter shows that SMM gives an accurate description
of the phase space experimentally 
explored~\cite{Bib_DAg3,pierre,Bib_Bou2,Bib_Bon1,gsi}.
These findings make the analysis of temperatures even more intriguing.

\section{Temperature of the QP source}
As we have shown the relevance of the statistical approach to 
describe the selected events, we will now present an analysis about 
thermodynamical observables of the multifragmenting source.
In particular we will extract from the data information about the freeze-out
temperature of the decaying system. This experimental  temperature 
will be compared with the predictions of two statistical models 
MMMC~\cite{grossreport90} and SMM~\cite{Bib_Bon2}, which use somewhat 
different assumptions on the freeze-out configurations. 

The first model~\cite{grossreport90} assumes that the break-up volume is 
fixed for all partitions. Primary fragments are produced in 
ground and low excited states and they de-excite only by evaporating neutrons.
Their de-excitation is treated in a microcanonical way inside the freeze-out
volume. 
In the second model~\cite{Bib_Bon2} one assumes that the break-up volume grows 
with the number of the fragments. 
Primary fragments have the same temperature as the whole system and 
in a later stage, during the Coulomb acceleration, they de-excite by 
evaporating neutrons and charged particles or by secondary Fermi 
break-up. 
In addition, at low excitation energies (typically up to 4-5 A.MeV),
some final fragments are sequentially emitted in the decay of the 
heavy residual nucleus present at freeze-out. 

From a general point of view, one should bear in mind that partitions at 
freeze-out are different from the asymptotic measured ones.
Therefore, following the two different freeze-out assumptions of the models 
we reconstructed primary partitions and we estimated the 
temperature $\Theta$ of the QP in each event looking for a new thermometer 
mainly based on fragments.
The energy balance at the freeze-out time reads:
\begin{equation}
m_0 + E^* = \left[ \sum m_i + \sum E_i^{int} + E_{coul} + \frac{3}{2} (M-1)
\Theta
\right].
\label{freeze}
\end{equation}
Here $m_0$ is the mass of QP, 
$E^*$ is the excitation energy of the source calculated via calorimetry 
through equation~(\ref{calorimetry}), $m_i$ $(i=1,\dots,M)$ 
are the masses of primary fragments, light particles and neutrons, 
$M$ is the total multiplicity, $E_i^{int} = a \Theta^2$ is 
the internal energy of primary fragments ($Z \ge 3$),
$E_{coul}$ is the Coulomb energy of the partition 
calculated by randomly positioning non overlapping
primary fragments in  the freeze-out volume.
The free parameters entering the equation at the freeze-out are the level 
density $a$ and the freeze-out density $\rho$. 

To evaluate for the data the different terms of eq.~(\ref{freeze}) 
we followed the two different assumptions of 
the models MMMC~\cite{grossreport90} and SMM~\cite{Bib_Bon2}.
To mimic the MMMC assumption we assumed for primary fragments ($Z \ge 3$) a 
charge-to-mass ratio as in the entrance channel. 
In the SMM hypothesis we shared among final fragments the
total charge detected in form of light particles and we assumed, as before,
a charge-to-mass ratio as in the entrance channel. 
This sharing was performed in two different approximations. In the first 
it was assumed that primary fragments will evaporate particles proportionally 
to their charge, while in the second approximation the same amount of charge 
(detected in form of light particles) was distributed to all the fragments 
to mimic a higher Fermi-break-up probability. These two assumptions give 
results in agreement within few percent.
In addition, following Ref.~\cite{Bib_Bon2} we allowed compound nucleus 
emission to simulate a secondary emission for the detected fragments.
We have bound the lightest fragment in each event with the largest one 
with a probability 
\begin{equation}
P(\epsilon^*) = e^{ -\frac{1}{2} ({\epsilon^* / \sigma})^2}.
\end{equation}
The width of the gaussian $P(\epsilon^*)$ was assumed to be $\sigma=4 \ A.MeV$,
based on the behaviour of the experimental fragment multiplicity as a function 
of the excitation energy. 
Indeed the two-fragment channel is experimentally seen in a
wide range of excitation energies, up to $\sim 4-5 \ A.MeV$, until the largest
fragment in each event becomes smaller than the Fission 
Fragments.

$\Theta$ defined algorithmically by equation (\ref{freeze}) 
has to be interpreted as an event by event estimator of the 
microcanonical temperature~\cite{Bib_Bon2,grossreport90}.
For each $E^*$, $\Theta$ has a spread according to the kinematical 
properties of the partitions. Similarly to SMM model, indeed, 
the {\it partition temperature} depends on the amount of kinetic energy still 
available after the subtraction of the Q-value to create a partition and of the
Coulomb energy of the interacting fragments.
Averaging $\Theta$ over all the partitions at fixed $E^*$ 
one gets an estimate of the microcanonical temperature of the system.

Equation~(\ref{freeze}) was first applied to the model events
to estimate how this backtracing from $t = \infty$ (final partitions) to 
$t = 0$ (freeze-out configuration) is able to approximate the original 
correlation between the microcanonical temperature and the excitation energy. 
The method resulted to work quite well as it reproduces the microcanonical
temperature within 10\% in the case of MMMC model and within 5\% in the SMM 
case.  Since equation~(\ref{freeze}) resulted reliable, hereafter we 
use the notation $T$ instead than $\langle \Theta \rangle$.

In Fig.~8 the experimental caloric curve, calculated through 
the total energy balance (eq.~(\ref{freeze})), is shown for the two freeze-out
hypotheses {\it SMM-like} and {\it MMMC-like}.
A freeze-out density $\rho = \rho_0/3$ has been assumed for the first
hypothesis on primary fragments and a density $\rho = \rho_0/6$ for the second
one.
Concerning the level density parameter we used $a = f(A)$, with $f$ 
continuously varying from $A/8$ for the lightest nucleus $A=7$ to $A/12$ for 
the heaviest one $A=197$ to take into account different surface and volume
contributions~\cite{Bib_Bon2}.   
In this figure we also report the caloric curves of SMM and MMMC models, 
calculated for filtered events.
The effect of the experimental acceptance on the caloric curve 
(an increase of the order of 15\% at the highest excitation energy)
is to suppress events with a large ($>7$) number of fragments which
correspond in the model to partition temperatures lower than those
corresponding to a limited number of fragments.
Therefore the expected plateau-like behaviour~\cite{grossreport90,bond98}
is no more evident.

Despite the difference in the absolute values of the temperature 
many similarities exist between the experimental $(T, \epsilon^*)$ 
reconstructed with the two hypotheses {\it SMM-like} and {\it MMMC-like}.
In both cases at small excitation energies 
$T \propto \sqrt{\epsilon^*}$ and becomes flatter 
for increasing energies even if the energy of this transition depends
on the freeze-out hypothesis.
If we consider the behaviour $T \propto \sqrt{\epsilon^*}$ 
as an indicator of a liquid phase, we can interpret this change 
as a first qualitative sign of a phase transition. 

To check the influence of the collective energy present in the experimental 
kinetic energies and of the parameters $a$ and $\rho$ used in 
equation~(\ref{freeze}), we have compared the experimental caloric curves 
obtained under different conditions.
We have found that any combination of parameters does not
modify significantly the experimental $(T, \epsilon^*)$ correlations,
the main change on the value of the freeze-out temperature
arising from the freeze-out reconstruction hypothesis.

It is interesting to note that the hypothesis {\it SMM-like} applied to the
data gives values of the temperature systematically higher than those given 
by the hypothesis {\it MMMC-like}.
This can be explained by the fact that, when applying eq.~(\ref{freeze}), we 
{\it always} or {\it never} bind the light charged particles to final 
fragments to get primary fragments.
Since however it is not experimentally established which is the excitation 
energy where a transition between the two regimes could take place  we can 
consider our experimental caloric curves as an upper and a lower limit to 
bracket the freeze-out temperature and we estimate the accuracy of our 
reconstruction to be about 2 MeV.
The contribution of secondary decay to the light charged particles and fragments
yield has therefore to be evaluated before a more precise determination
of the freeze-out temperature can be done.

The fact that SMM and MMMC for a given excitation energy and density
produce approximately the same asymptotic partitions~\cite{sneppen,pierreprep}
suggests that 
static variables do not allow to discriminate between the two 
freeze-out hypotheses. Therefore we investigated dynamical observables
that could keep a memory of the different fragment production mechanisms, such 
as for instance the correlation functions of the reduced velocity:
$v_{red}  = \frac{ \mid \vec v_i - \vec v_j \mid} {\sqrt (Z_i+Z_j)}$.

It is well established~\cite{Bib_DAg4,correl} that at small values of 
the relative velocity the two-fragment correlation functions of the reduced
velocity are quite sensitive to the interfragment emission time and to the
density of the emitting system.
Here we want to exploit another property of this kinematical region,
called {\it Coulomb hole}.
Namely a partial filling of the Coulomb hole would reveal a decrease of 
correlation among fragments due to their late decay through light charged 
particle emission.

We show in Fig.~9 the correlation functions of the reduced velocity
for fragments with charge $4 \le Z \le 15$ for theoretically calculated 
events at excitation energies $4 \ A.MeV$ and $6 \ A.MeV$. In the upper panel 
MMMC predictions are presented and in the lower panel SMM events are analyzed.
In this case the codes were run at the same density ($\rho_0/6$) to enhance
in the Coulomb hole region only effects due to secondary evaporation.

The sensitivity of this dynamical observable to the different hypotheses of
fragment production is evident. We clearly see how the SMM assumptions about 
Fermi break-up and light particle evaporation decreases the correlation
among fragments with respect to the MMMC assumption of fragments which freeze 
only through neutron emission. 

Since the statistics of the measured events presented in this paper is too low
to perform a calculation of correlation functions in different bins of the
excitation energy, we present in Fig.~10 only two large intervals of 
$\epsilon^*$ which should be representative of a possible change in the 
fragment production mechanism.
In the upper panel of Fig.~10 the correlation functions for 
$\epsilon^* < 3.5 \ A.MeV$ are shown and in the lower panel the same observable
for $5 \le \epsilon^* \le 7 \ A.MeV$.
Both panels refer to fragments with charge $4 \le Z \le 15$.

Even if the statistics is poor, the difference between the two
correlation functions is evident. The width of the Coulomb hole increases
with increasing $\epsilon^*$ indicating a transition to a prompter emission 
of fragments. 

The experimental correlation functions of the reduced velocity cannot be 
directly compared with the predicted ones presented in Fig.~9.
Indeed the model calculations are done at a fixed excitation energy, while in
the data the observables result from an average of events with 
different excitation energies. In addition an investigation about the
density of the decaying system should be performed before comparing dynamic
observables~\cite{pierre}, but this is beyond the aim of this paper.
We can only claim that data show a lower correlation among fragments 
than the one seen in MMMC model. 
The hypothesis of hot fragments with an important secondary emission 
seems more reliable, at least for our dataset. 
We should note that in Ref.~\cite{nmarie} a sophisticated 
correlation analysis suggests the existence, in the multifragmentation of 
highly excited sources, of hot primary fragments
emitting light charged particles.

More statistics is needed to perform a quantitative comparison between 
experimental and predicted correlation functions in bins of excitation energy.
We hope that measurements with high statistics will be soon performed by new 
$4\pi$ detectors~\footnote{For instance the detector Chimera~\cite{chimera}, 
which will start to be operating in 1999 at the Superconducting Cyclotron of 
the Laboratorio Nazionale del Sud, 
Catania} with high mass and energy resolution, with the aim of studying, for 
a well defined source, the transition from sequential to prompt fragment 
emission through the correlation functions of the reduced velocity.

To summarize this Section, we have shown that different freeze-out assumptions
lead to different back-traced freeze-out temperatures and thus to a 
different origin of the measured partitions and of the isotopic yields.
From correlation functions we got the indication that dynamic observables could 
be better reproduced by models where "hot" primary fragments are generated
at the freeze-out, decaying later through light particle and fragment emission.
In our range of excitation energies the isotopic temperature
seems to be a measure of the mean temperature of the emitting systems at a 
stage of the reaction later than the freeze-out.

\section{Signals of phase transition}
\subsection{Fluctuations}
After the study of average temperatures, we analyzed fluctuations of 
($\Theta,\epsilon^*$). It is hopeful indeed that $\Theta$ fluctuations 
from event to event will provide a deeper understanding of the thermodynamical 
properties of the decaying systems.
In Fig.~11 we show the ($\Theta,\epsilon^*$) scatter plots for data and SMM
events, calculated in a larger energy range.
It is clearly seen that in both cases the distribution has a broadening
in the region around $4 \div 6 \ A.MeV$.

One of the observables related to energy fluctuations is the specific heat
capacity.
A peak for this observable at the phase transition was anticipated by 
theoreticians more than 10 years ago~\cite{grossreport90,Bib_Bon3} and
more recently studied in the framework of different 
models~\cite{cv,mishustin,gulm_phil2}.

For a canonical ensemble the specific heat capacity at constant volume can be
calculated as:
\begin{equation}
C_V = \frac{\sigma^2_E}{T^2} = 
\frac{ \langle E^2 \rangle - \langle E \rangle ^2} {T^2}.
\label{cv}
\end{equation}
Since it is expected that $C_V$ shows a peak at the phase transition 
temperature, this gives a hint that also in our case $\sigma_E^2/T^2$ may
be relevant to provide information in addition to the average
characteristics.
To better investigate this point, we study the energy fluctuations
extracted from data on an event by event basis. 

We should nevertheless emphasize that in order to study energy variances, in 
our microcanonical type of framework, we have to consider fixed bins of 
$\Theta$ and then combine various partitions with different energies. In this 
way we deal with energy fluctuations even if these fluctuations do not 
strictly coincide with those of a canonical ensemble.
The absolute value of the total energy variance $\sigma_E^2/T^2$ cannot 
in our case be directly interpreted as a heat capacity but it will rather
be related in a non linear way both to the heat capacity and the latent heat 
of the transition.
In any case even for a microcanonical system
the location of the peak, if any, will sign the occurrence of a
phase transition~\cite{gulm_phil2}.
                   
In Fig.~12 we plot the experimental $\sigma_E^2/T^2$ as a 
function of the temperature for the two freeze-out hypotheses
and for the different values of the parameter $a$ 
used in the calculations. We show also calculations where 
the extra-kinetic energy was subtracted or included in the fragment kinetic
energies.
The dashed lines in Fig.~12 correspond to regions where the statistical 
error of the events in the bin of temperature is larger than 10\%. In all the 
cases $\sigma_E^2/T^2$ shows a peak and this qualitative behaviour is not 
modified if we vary the parameters $\rho,a$ or we include the extra-kinetic 
energy.

One may worry if the decrease of fluctuations at high temperature
is not trivially induced by the shape of the experimental total
energy distribution: a monotonically increasing function could be biased
by the decreasing weight of high excitation energies.
To explore this possibility we have artificially flattened 
the excitation energy distribution giving to the detected events 
an energy dependent weight for the calculation of eq.~(\ref{cv}).
The result is a deformation of the shape of the fluctuations without any 
shift in the location of the peak.
To complete this analysis and to verify if experimental inefficiencies 
and limitation at high excitation energy could modify the distribution of 
this observable, we report in Fig.~13 $\sigma_E^2/T^2$ calculated for 
SMM not filtered events under different conditions.
The lack of events at high excitation energies (temperatures) modifies 
the upper tail of the distribution and reduces the height of the maximum 
but again does not produce any shift in the position of the peak which appears
to be robust.

We can then conclude that the observation of a peak in the experimental 
effective specific heat, defined by eq.~(\ref{cv}) and shown in Fig.~12,
reveals a phase transition for the studied system.
The two different freeze-out hypotheses 
lead to transition temperatures differing by about 1.5 MeV but almost identical
excitation energies.
However, one should remember that the ensemble of experimental events is 
limited in excitation energy. Data at a higher beam energy, leading 
to a wider excitation energy distribution, would be most welcome to confirm 
this finding.

Another way to characterize fluctuations is to analyze the behaviour of 
($\Theta,\epsilon^*$) for different fragment partitions. 
One of the most powerful methods to characterize 
the critical behaviour of a system undergoing multifragmentation is the 
method of conditional moments introduced by Campi~\cite{campi}.
The moments of asymptotic cluster charge distributions are defined as:
\begin{equation}
m^{(j)}_{k} = \sum_{Z} Z^{k} n^{(j)}(Z)
\label{mm}
\end{equation}
and the normalized moments as:
\begin{equation}
S^{(j)}_{k} = m^{(j)}_{k}/ m^{(j)}_{1} 
\label{ss}
\end{equation}
where $n^{(j)}(Z)$ is the multiplicity of clusters of charge $Z$ in the event 
$j$, and the summation is over all the fragments in the event {\it except the
heaviest one} which corresponds to the bulk liquid in an infinite system.

Fig.~14~a) shows, for each event $j$, the experimental scatter plot
of the logarithm of the charge of the largest fragment ($ln(Z^{(j)}_{big})$) 
as a function of the logarithm of the corresponding second moment 
($ln(m^{(j)}_{2})$) (Campi scatter plot). If the system experiences a
phase transition this plot should exhibit two branches: an upper branch with an
average negative slope corresponding to under-critical events and a lower
branch with a positive slope that corresponds to over-critical events.
The two branches should meet close to the critical point of the phase transition
~\cite{cmdtheo,gross1,campi}. 

As already done in previous studies of peripheral 
collisions~\cite{letter,subs}, to characterize the two branches and the 
intermediate zone, we have made three cuts in this plot selecting the
upper branch (Cut 1), the lower branch (Cut 3) and the
central region (Cut 2) and analysed the events falling in each of the
three zones.
It was already shown~\cite{letter,subs} that the fragment charge 
distributions obtained in the three zones exhibit shapes going from a "U" 
shape in Cut 1, characteristic of evaporation events, to an exponential 
one for Cut 3, characteristic of highly excited systems. In Cut 2 a 
power law fragment charge distribution $Z^{-\tau}$ 
($\tau \approx 2.2$) was observed as expected according to the 
Fisher's droplet model for fragment formation 
near the critical point of a liquid-gas phase transition~\cite{fisher}.

Here we argue that the Campi plot can be used also for a calorimetric analysis,
to study the thermodynamical properties of the emitting Quasi-Projectiles 
for events belonging to the different cuts even if each selected region is 
representative but not exhaustive of a particular class of events.

The upper branch contains events having an average reduced
impact parameter $0.98$, mean excitation energy $1.4 \ A.MeV$, while 
the lower branch consists of events having 
$\langle b/b_{max} \rangle = 0.67$, 
$\langle \epsilon^* \rangle = 6.4 \ A.MeV$. The region where the two branches 
meet is populated by events with $\langle b/b_{max} \rangle = 0.81$, 
$\langle \epsilon^* \rangle = 4.3 \ A.MeV$.
Remark that $b/b_{max} \approx 0.8$ was found in theoretical calculations
based on Classical Molecular Dynamics~\cite{theomoh} as the value of 
the impact parameter where the QP shows the signals of a critical behaviour.
The temperature distributions of the QP depend on the assumptions 
made for the freeze-out configuration as discussed in the previous Section.
Following the hypothesis where both neutrons and light charged particles are 
emitted by primary fragments we obtain for the cuts 1, 2, 3 on the Campi 
scatter plot mean partition temperatures of $3.9, 5.8, 6.4 \ MeV$, respectively.
Following the hypothesis of fragment de-excitation only through neutron 
emission we obtain for the three cuts mean partition temperatures 
$2.9, 4.0, 4.2 \ MeV$, respectively.

The contributions to the caloric curve in the three cuts are shown by different 
colours in Fig.14~b) for experimental events. For comparison in Fig 14~c),~d) 
we show the Campi plot and the caloric curve for SMM filtered events.
It is clearly seen, both for data and SMM events, that for cuts 1 and 3 there 
are unusual events (although with low probability) which lie far from the 
average $T(\epsilon^*)$ behaviour. These are unusual compound-like
states with very large temperatures and multi-fragments events with
low temperatures.
For these events one could do the analogy respectively with an overheated liquid
and a super-cooled gas for ordinary liquid-gas phase transition.
Their yield in the present experiment is too small to draw 
reliable quantitative conclusions.
Further studies on $\Theta$ fluctuations will provide a deeper 
understanding~\cite{gulm_phil2} of the thermodynamical 
properties of the under-critical, critical and over-critical 
events of the phase transition in finite systems.

\subsection{Critical exponents}
If multifragmentation is a manifestation of a phase transition 
in nuclear matter one can try to extract from the data a set of 
critical exponents characterizing this phase transition. Strictly 
speaking critical exponents are defined for 2-nd order phase 
transitions where the correlation length diverges. Indeed
a liquid-gas phase transition is 2-nd order only at the 
critical point. Nevertheless, as we have seen in Fig.~11, a large transition 
region with wide energy fluctuations is populated by the selected 
multifragmentation events.
One may hope that some events fall in the vicinity of the
critical point.
On the other hand in finite systems the universal critical behaviour 
can be also induced by finite size effects (see e.g. lattice based 
calculations~\cite{Gulm_phil}).

For systems undergoing a second order phase transition the moments of the 
charge distribution (eq.s~(\ref{mm}) and (\ref{ss})) $m_{k}$ and $S_{k}$ should 
reveal some specific correlations~\cite{campi} and in particular for $k \ge 2$
they should diverge at the critical point or show a peak, due to finite size 
effects. In previous papers~\cite{letter,subs,rassegna} peaks were found for
the second moment $m_2$, the reduced variance $\gamma_2$ and the normalized 
variance of the largest fragment $\sigma_{NV}$ as a function of the charge 
particle multiplicity. These are indications that a critical behaviour 
could occur.

Critical exponents~\cite{eos,elliot98} are characteristic of the specific 
universality class of the transition.
Here we recall that the exponent $\tau$ gives the slope of the power-law
distribution at the critical point, the exponent $\beta$ fixes the behaviour 
of the largest cluster close to the critical point and finally the exponent
$\gamma$ governs the strength of the singularity of $m_2$ at this point.
These quantities are connected by universal relations~\cite{fisher,theo}.

We could not extract $\gamma$ from the second moment of the
charge distribution, as performed in Ref.s~\cite{eos,elliot98}, because in our
range of excitation energies we have too few events in the over-critical region
to constrain the value of $\gamma$. 
Instead of this the ratio $\beta / \gamma$ has been calculated by fitting the 
correlation $ln(Z_{big})$ versus $ln(\langle S_{2} \rangle)$, mainly in the
"liquid" region. The value of $\tau$ has been obtained from the correlation 
between moments with different order 
$S^{j}_{k}$ and $S^{j}_{l}$ ($l > k, k \ge 2$), since the slope of the 
log-log correlation is $(\tau-l-1)/(\tau-k-1)$. 
The exponent $\tau$ has been also calculated
by fitting the Z-distribution in the "critical" region (Cut 2 in the 
Campi-scatter plot).
The exponent $\beta$ has been obtained by the fit of 
the mean value of the charge of the heaviest fragment $Z_{big}$ as a function 
of the excitation energy (temperature), assumed to be the critical parameter. 

Before calculating $\beta / \gamma$ from $ln(Z_{big})$ versus 
$ln(\langle S_{2} \rangle )$, great care was payed to the elimination of 
fission events, following the warnings given in Ref.~\cite{gross1}. 
These events indeed, enhancing the value of the moments, do not
allow a proper determination of the critical exponents.
From this study it resulted that there are two regions not affected by the 
criterion chosen to eliminate fission events: the higher part of the 
{\it liquid} branch and the lower branch. These regions have been considered to 
determine $\beta / \gamma$.
By fitting (Fig.~15~a) the upper part of the correlation we obtain 
$\beta / \gamma = 0.29 \pm 0.01$, while from the lower branch we get 
$1 + \beta / \gamma = 1.31 \pm 0.07$.
In the latter case the error is large, because the fitted region 
in our case is limited (we do not observe vaporization events).

We have calculated the ratio $\beta / \gamma$ also for SMM predictions.
For not filtered events we get $\beta / \gamma = 0.28 \pm 0.01$ for the upper 
branch while from the lower branch, where at high 
excitation energy the model gives mainly vapor, we obtain
$1 + \beta / \gamma = 1.28 \pm 0.02$.
For filtered events we obtain $\beta / \gamma = 0.28 \pm 0.02$ and
$1 + \beta / \gamma = 1.28 \pm 0.03$.
It is important to note that the fits of the two branches give the same values 
of $\beta / \gamma$ even in the case of filtered events and that 
these values are in agreement with those found from data.
The theoretical value expected for a liquid-gas phase transition is
$\beta/\gamma=0.265 \pm 0.006$.

The exponent $\tau$ has been calculated from the correlations 
$S_3$ - $S_2$ and $S_4$ - $S_3$.
In the case of SMM events the fit of $S_3-S_2$ gives $\tau=2.168\pm0.002$.
Experimentally, by fitting $S_3$ versus $S_2$ (Fig.~15~b)) we obtain 
$\tau = 2.12 \pm 0.02$. 
To verify the extent to which the experimental inefficiencies for light particle
detection influence the value of $\tau$ we have calculated the same 
correlation by correcting for the mean efficiency of the apparatus to the 
contribution of the light particles. With this correction we get
$\tau= 2.13 \pm 0.02$ compatible with the previous value.
By fitting higher moments, like $S_{4}$ versus $S_{3}$, we obtain 
$\tau = 2.13 \pm 0.01$.

In Fig.~15~c) we report the correlation $S_3 - S_2$ for the events falling in 
the regions 1, 2 and 3 of the Campi scatter plot (respectively squares, 
points, open circles).
We observe that the events falling in the region 2 cover the 
whole range of the $S_3 - S_2$ correlation thus determining $\tau$. 
Evaporation events belonging to the Cut 1 fall in the 
region of small values of $S_{3}$ and $S_{2}$ and 
multifragmentation events belonging to the Cut 3 fall in the 
upper corner of the correlation.

We show in Fig.~15~d) the charge distribution for events belonging to the
Cut 2 of the Campi-scatter plot. The line represents the power-law fit of 
the distribution which gives $\tau = 2.13 \pm 0.08$, in agreement 
with the value $2.22 \pm 0.01$ obtained for $\tau$ from the relation 
$\tau = 2+ 1/(1+\gamma/\beta)$.
It was claimed by several theoretical works~\cite{campilattice,richert}
that it is difficult to compare the exponents obtained for finite systems
with the expected ones corresponding to the relevant universality class, due 
to the deformations induced by finite size effects.
However the slope $\beta / \gamma$~\cite{gross1}, as well as the related 
exponent $\tau$~\cite{Gulm_phil}, rapidly approach the value 
expected in the thermodynamic limit for increasing size of the system.

As previously mentioned, to extract from data another critical exponent,
we need to make an assumption about the critical parameter. It is predicted 
~\cite{gross1,campi} that $Z_{big}$, the bulk in the liquid region, behaves as:
\begin{equation}
Z_{big} \sim \epsilon ^{\beta} 
\label{beta}
\end{equation}
where $\epsilon \ (>0)$ is a variable characterizing the vicinity to the 
critical point. 
In percolation $\epsilon =  p - p_c \ (p>p_c)$ is the deviation from the 
threshold probability $p_c$, while in thermal transitions of the liquid-gas 
type in the canonical ensemble 
$\epsilon = T_c - T \ (T<T_c)$ is the deviation from the critical 
temperature.

Since experimentally $\epsilon$ is not directly observable, a monotonic 
function of $\epsilon$ can allow to determine critical exponents. For instance
the multiplicity of produced cluster~\cite{eos,campi,elliot98} has been often
used.
An alternative possibility is the thermal excitation energy $\epsilon^*$
which could be the proper thermal critical parameter in the microcanonical 
ensemble~\cite{gross1,quattro}. This observable, together with the estimated 
temperature extracted from equation~(\ref{freeze}), has been used in this work 
for the determination of the exponent $\beta$.

In Fig.~16~a) the correlation $Z_{big}- \epsilon^*$ is shown,
together with the fit resulting from equation~(\ref{beta}).
From this fit we obtain $\epsilon^*_{crit} = 4.5 \pm 0.2 \ A.MeV$
and $\beta = 0.33 \pm 0.04$.
We want to emphasize that in the range of excitation energy used for the fit 
($1 - \ 4 \ A.MeV$), the possible contribution of collective energy to the 
thermal excitation one is negligible so that the value of $\epsilon^*_{crit}$
is actually reliable.
In Fig.~16~b) we show another representation of the behaviour of the size 
of the largest fragment as a function of the critical parameter:
$ln(Z_{big})$ vs. $ln(\epsilon^*_{crit}-\epsilon^*)$. This representation
should enhance the deviations from the power law behaviour (eq.(~\ref{beta})).

If instead of $\epsilon^*$ we use $\sqrt{\epsilon^*}$ as critical parameter
(which roughly corresponds to a temperature in the liquid side)
we obtain $\epsilon^*_{crit} = 5.0 \pm 0.3 \ A.MeV$ and $\beta = 0.33 \pm 0.02$.
By using the reconstructed temperature as critical parameter (Fig.~16~c) and d))
we get $T_{crit} = 6.0 \pm 0.4 MeV$, $\beta = 0.35 \pm 0.07$ with the 
freeze-out hypothesis SMM-like and $T_{crit} = 4.6 \pm 0.3 \ MeV$, 
$\beta = 0.30 \pm 0.07$ in the MMMC-like hypothesis.

The value of $\beta$ extracted from SMM predictions is 
$\beta = 0.344 \pm 0.004$, in agreement with experimental data.

The fact that a power law behaviour with the same 
$\beta$ exponent is found for $Z_{big}$ when 
working with $\epsilon^*$ or $\sqrt {\epsilon^*}$ or $T$ should not 
be surprising since this behaviour is expected 
only in the close proximity of the transition point.  
However this result confirms the fully statistical nature of the data,
the coherence of our analysis and the reliability
of eq.~(\ref{freeze}) to calculate the freeze-out temperature.

In conclusion we have shown that the correlation between the 
largest cluster and the excitation energy or the temperature leads to values 
of the critical parameter fulfilling our hypothesis on the 
relationship between $\epsilon^*$ and $T$.
The values of $\epsilon^*_{crit}$ and $T_{crit}$ correspond to the mean values 
obtained in the region 2 of the Campi-scatter plot. These values correspond 
to the transition region estimated from the analysis of the fluctuations,
where the effective specific heat shows a peak.
The values of $\beta/\gamma$, $\beta$ and $\tau$
are compatible with theoretical predictions for systems undergoing 
a phase transition.

We want however to remark that the extraction of critical exponents has
been mainly based on experimental correlations related to the {\it liquid} 
branch of the Campi scatter plot.
We think that an analysis similar to the one presented in this paper,
performed also for the {\it over-critical} branch of the Campi scatter 
plot, could give a more complete understanding of the nuclear phase transition.

\section{Summary and Conclusions}
In this paper we have presented a comprehensive study on the Quasi-Projectile
multifragmentation in peripheral collisions for the reaction Au + Au at 
35 A.MeV incident energy.
The data selection constraints and the source reconstruction technique
have been chosen to minimize preequilibrium and/or neck contamination
in the predominantly binary events detected.

The high degree of equilibration attained by the decaying QP system 
is also demonstrated by the agreement with statistical multifragmentation model 
predictions.
The excitation energy and temperature of the source have been evaluated through 
a detailed energy balance at the freeze-out time.
On an event by event basis the correlation between
the temperature and the excitation energy of the system has been calculated 
by backtracing the experimental information to the freeze-out time.

From an experimental point of view it has been shown the relevant importance of 
the fluctuations to characterize the multifragmentation phenomenon.
A peak of the "canonical" heat capacity indicates that the system undergoes 
a liquid-gas type phase transition at an excitation energy $\sim 5 \ A.MeV$
and a critical temperature $4 - 6 \ MeV$, dependent on the freeze-out 
hypothesis.

Critical exponents, extracted from the moments of the charge distribution
in the liquid part of the experimental Campi plot, 
have values compatible with theoretical predictions for systems undergoing 
a phase transition.

\section{Acknowledgements}
The authors would like to thank Philippe Chomaz and Francesco Cannata for many 
interesting, stimulating and constructive discussions.

The authors would like to acknowledge the {\it Multics-Miniball} collaboration, 
which has performed the $Au + Au$ experiment at $35 \ A.MeV$.

This work has been partially supported by NATO grants CRG 971512.
I.N.M. acknowledges support from Carlsberg Foundation (Denmark).

\newpage
\begin{figure}[h]        
\begin{center}
\epsfig{file=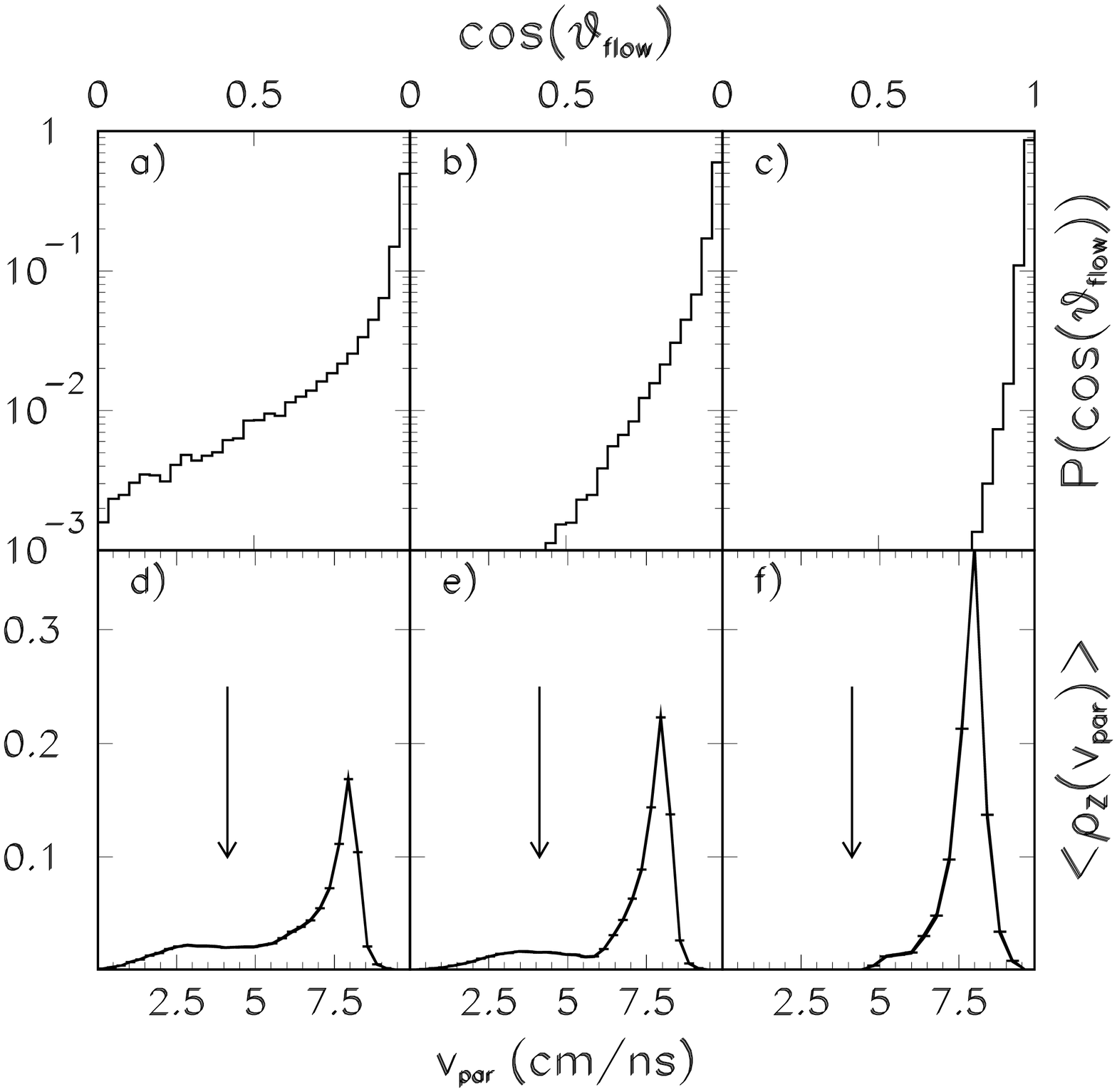,width=0.75\textwidth}
\end{center}
\end{figure}
\noindent
{\bf Fig. 1:} \\
Probability distribution of $cos(\theta_{flow})$ (upper panels)
and distribution of {\it charge density} $\langle \rho_Z (v_{par}) \rangle$ 
(lower panels) as a function of the fragment velocity along an axis
parallel to the QP velocity.
The arrow indicates the mid-rapidity velocity.\\
Panels d), e) and f) correspond to the same selections performed
for the panels a), ~b) and ~c), respectively (see text).
\newpage
\begin{figure}[h]        
\begin{center}
\epsfig{file=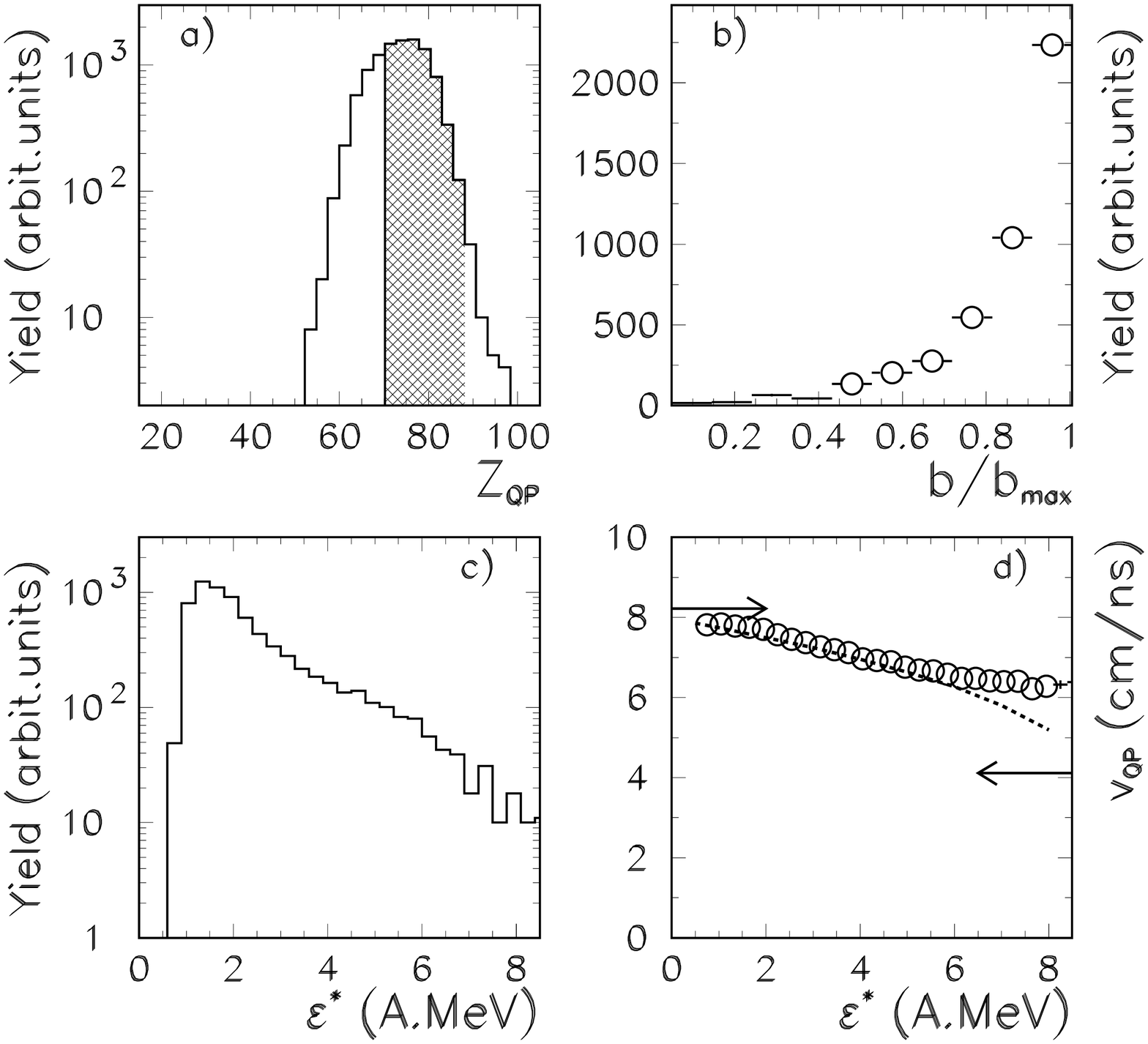,width=0.75\textwidth}
\end{center}
\end{figure}
\noindent
{\bf Fig. 2:} \\
a) Charge distribution of the $QP$.
The shaded area ($70 < Z_{QP} < 88$) represents the events analyzed 
in the present work.\\
b) Yield of the reduced impact parameter.\\
c) Experimental distribution of the $QP$ excitation energy, obtained with 
eq.~(\ref{calorimetry}).\\
d) Experimental correlation between the laboratory
velocity of the QP and its excitation energy.
The dashed line represents the values obtained from eq.~(\ref{conservation}).\\
Panels ~b), ~c) and ~d) refer to events belonging to the the shaded area of 
panel a).
\newpage
\begin{figure}[h]        
\begin{center}
\epsfig{file=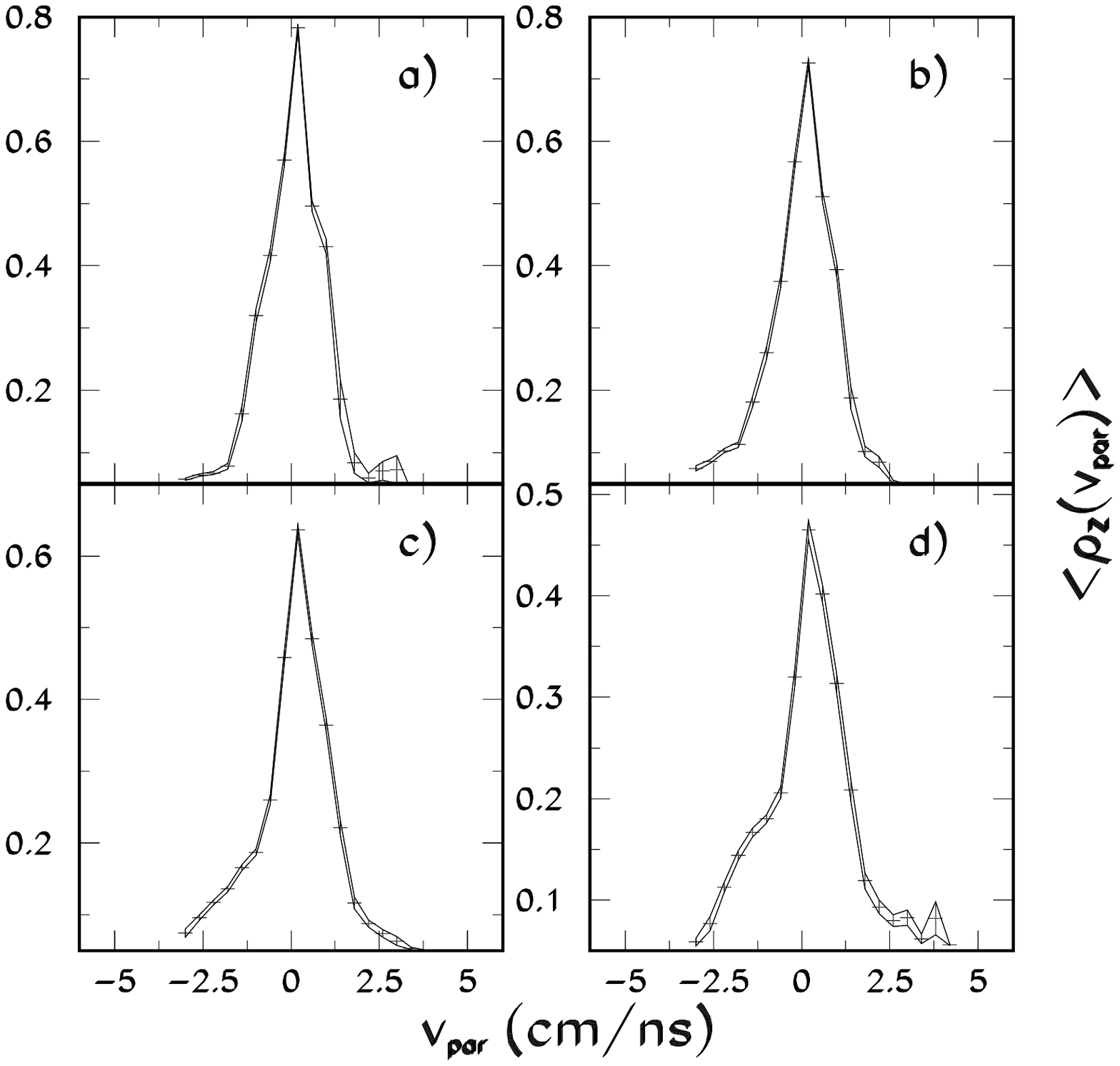,width=0.75\textwidth}
\end{center}
\end{figure}
\noindent
{\bf Fig. 3:} \\
{\it Charge density} $\langle \rho_Z (v_{par}) \rangle$ 
distribution as a function of the fragment velocity, along the 
axis parallel to the QP velocity.
Panels a), b), c) and d) refer to $\epsilon^* = 1 - 2 \ A.MeV$, 
$\epsilon^* = 2 - 3 \ A.MeV$, $\epsilon^* = 3 - 6 \ A.MeV$ and 
$\epsilon^* = 6 - 8 \ A.MeV$, respectively.
The mid-rapidity source is centered at $\sim -4 \ cm/ns$.
\newpage
\begin{figure}[h]        
\begin{center}
\epsfig{file=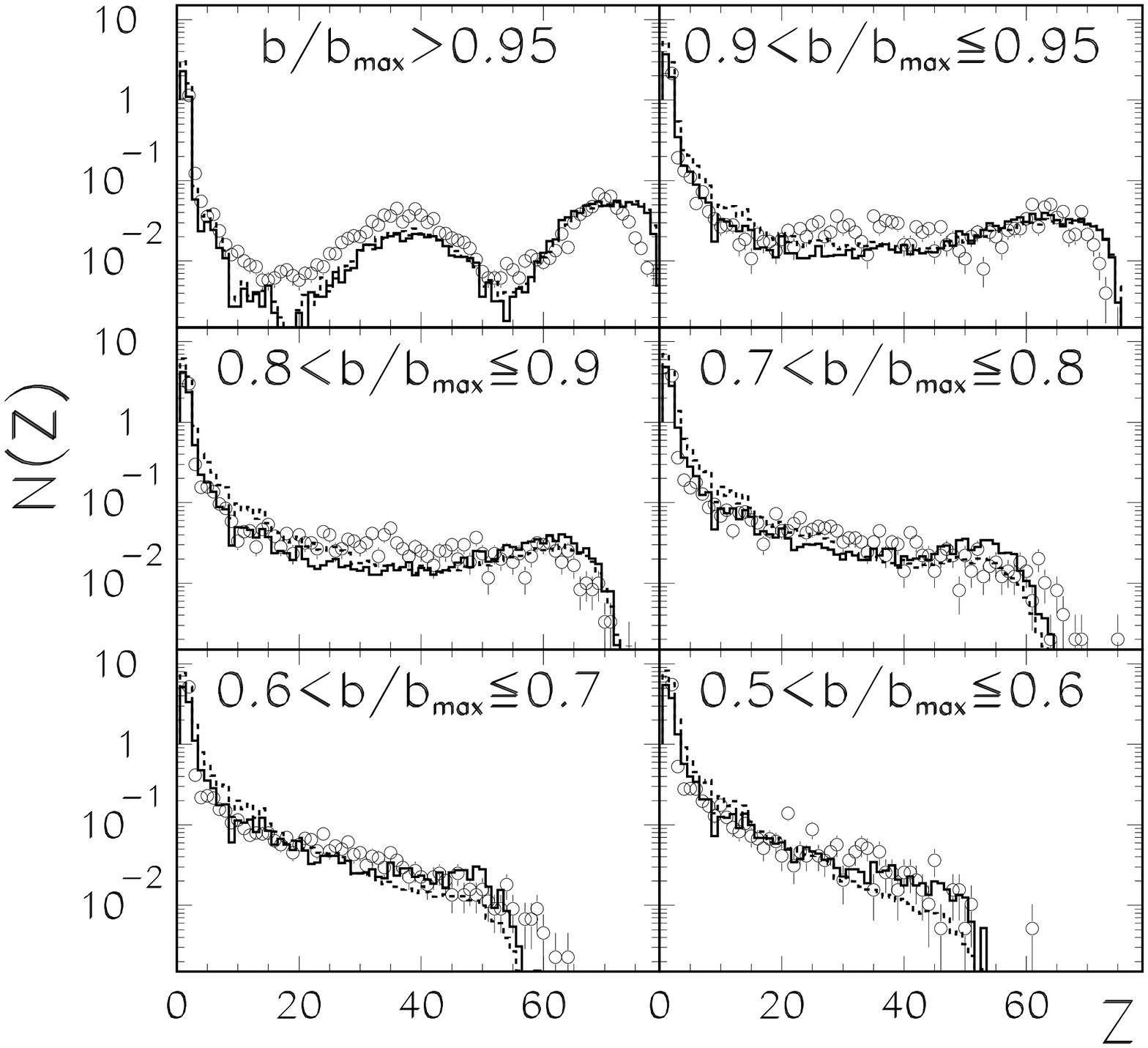,width=0.75\textwidth}
\end{center}
\end{figure}
\noindent
{\bf Fig. 4:} \\
Mean elemental event multiplicity $N(Z)$ in different $b/b_{max}$ intervals.
The circles show experimental data and the solid/dashed histograms the results 
of SMM filtered/not filtered predictions.
\newpage
\begin{figure}[h]        
\begin{center}
\epsfig{file=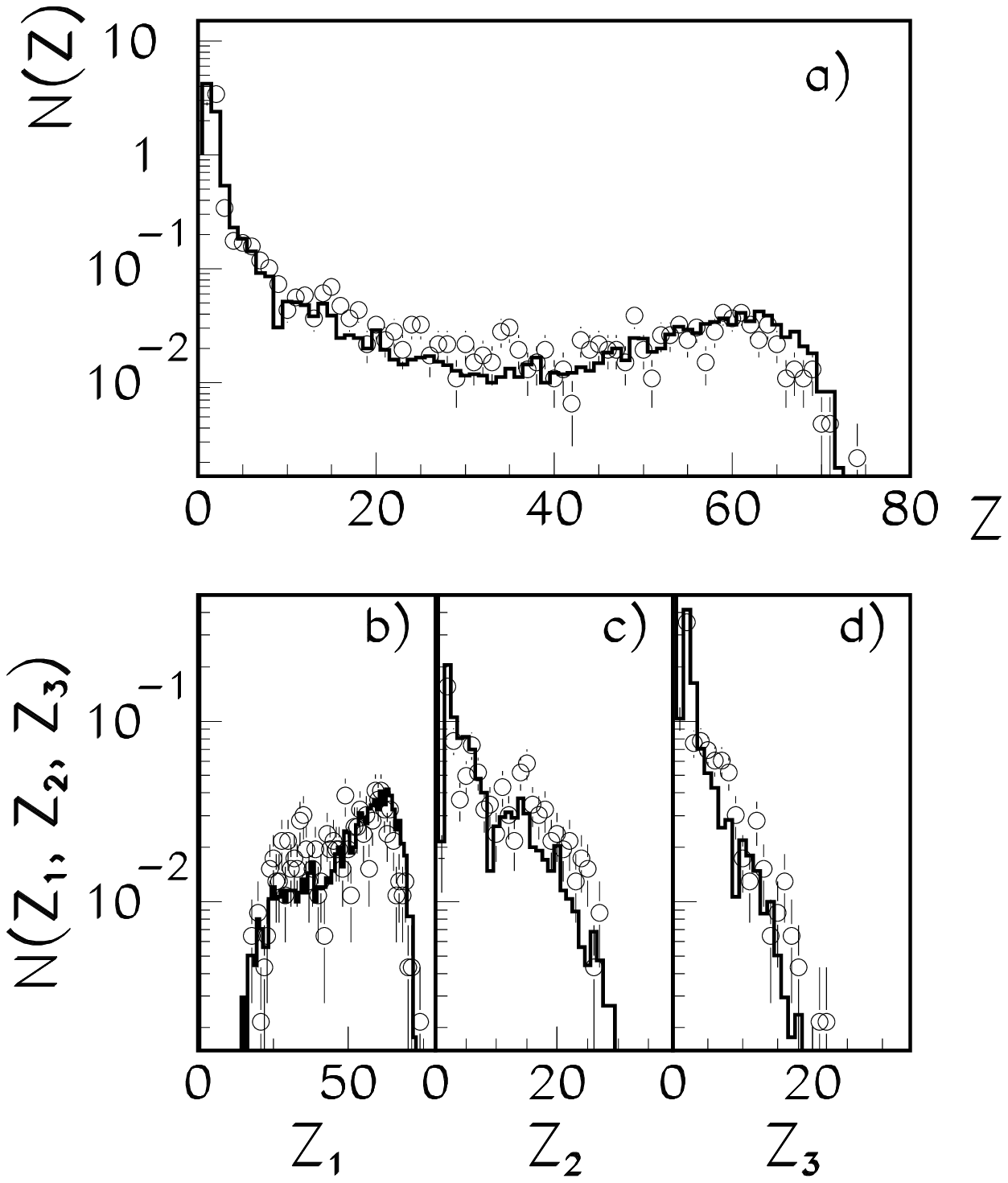,width=0.75\textwidth}
\end{center}
\end{figure}
\noindent
{\bf Fig. 5:} \\
Mean elemental event multiplicity $N(Z)$ and distribution of the three 
heaviest fragments in each event, ordered for decreasing size 
($Z_1 \ge Z_2 \ge Z_3$) for $0.8 < b/b_{max} \le 0.9$ with fission events 
removed.
The circles show experimental data and the lines the filtered SMM predictions.
\newpage
\begin{figure}[h]        
\begin{center}
\epsfig{file=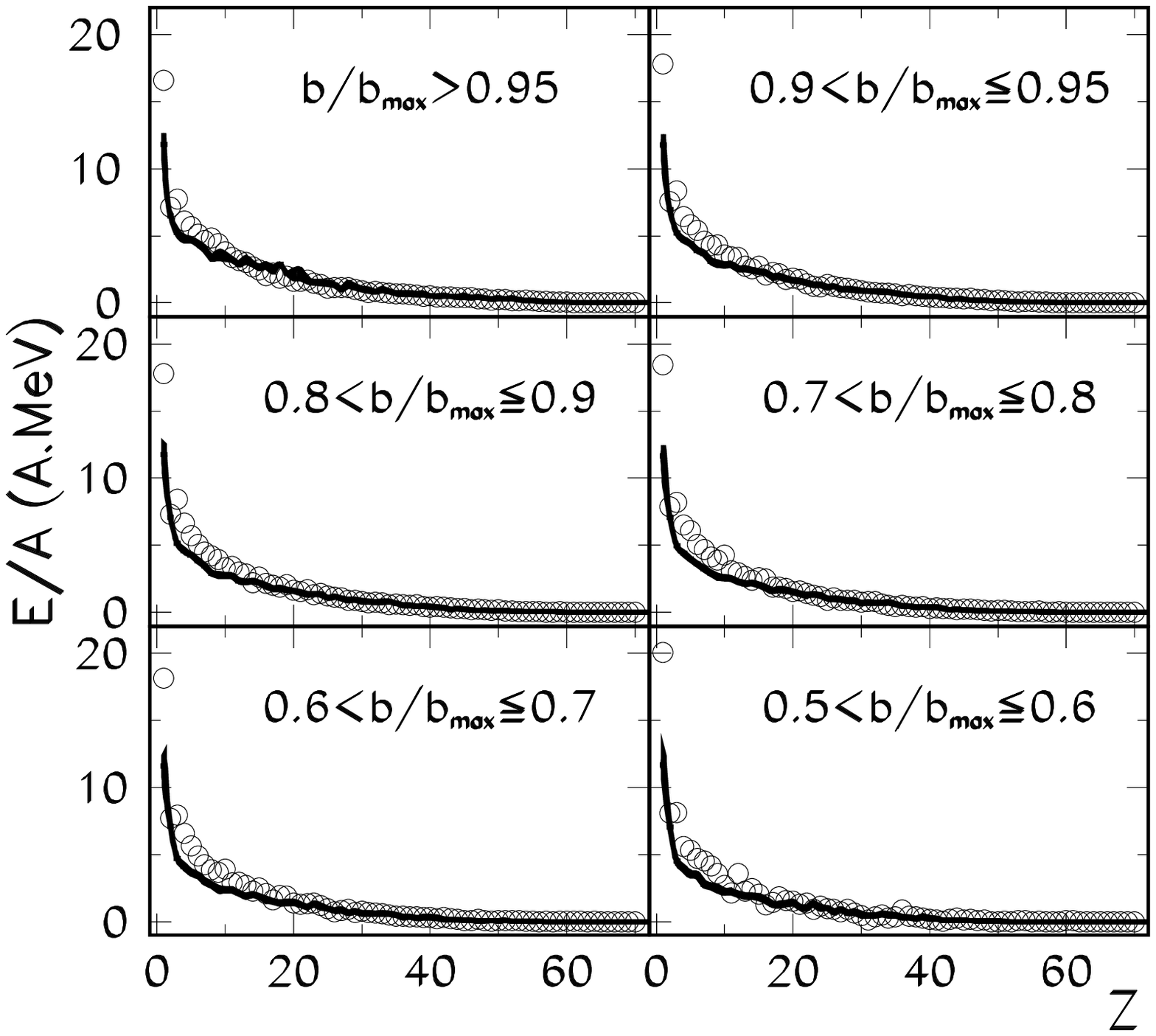,width=0.85\textwidth}
\end{center}
\end{figure}
\noindent
{\bf Fig. 6:} \\
Measured (circles) and predicted (lines) mean kinetic energy per 
nucleon as a function of the fragment charge for different $b/b_{max}$ 
intervals.
\newpage
\begin{figure}[h]        
\begin{center}
\epsfig{file=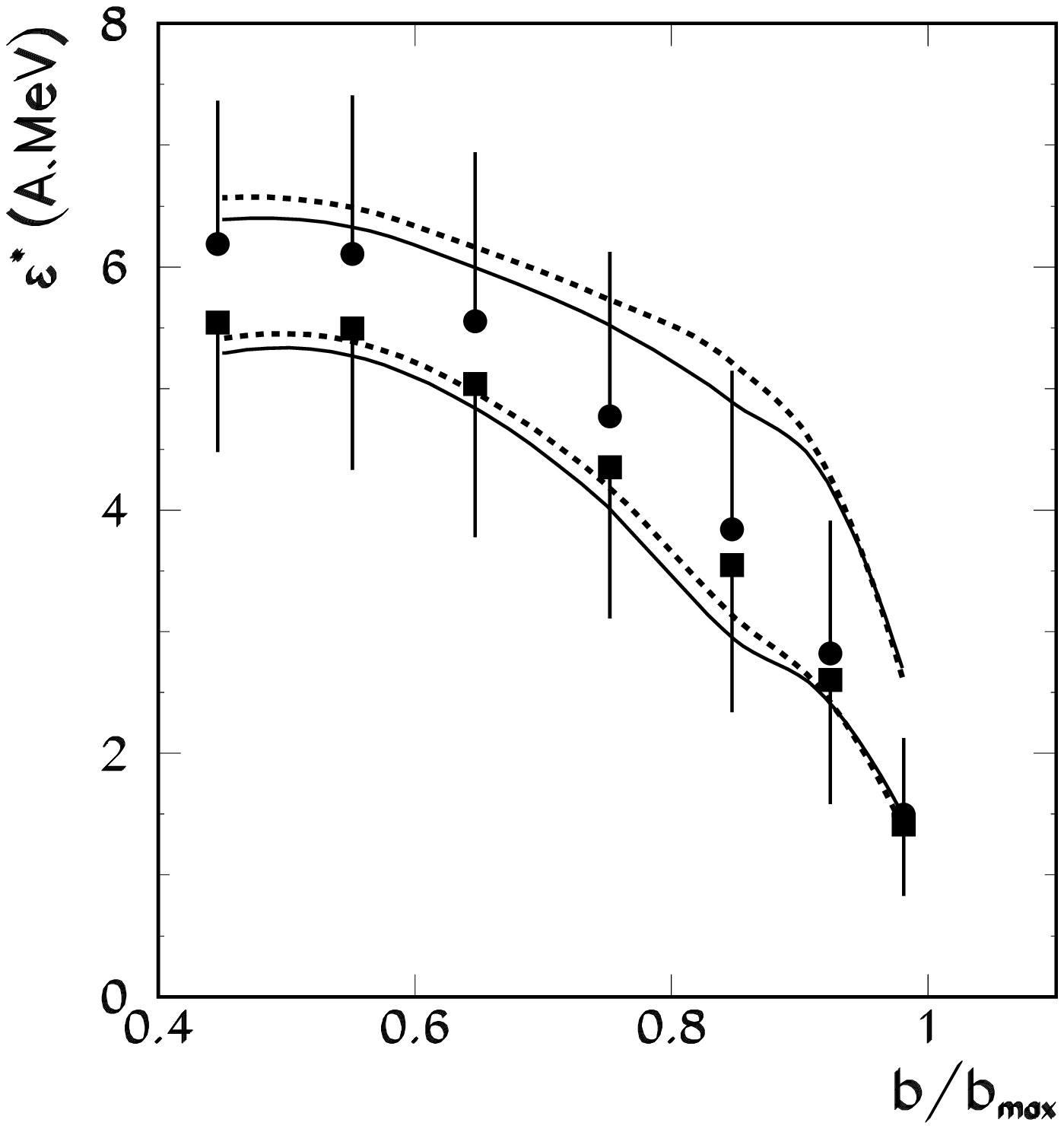,width=0.75\textwidth}
\end{center}
\end{figure}
\noindent
{\bf Fig. 7:} \\
Mean excitation energy per nucleon and its standard deviation as a function 
of the reduced impact parameter. 
Points show the experimental mean values (+~the standard deviation),
squares are the excitation energy decreased by the collective component
(-~the standard deviation).
Solid/dotted lines are filtered/not filtered SMM mean values $\pm$ the 
standard deviation.
Fission events were eliminated from both data and predictions.
\newpage
\begin{figure}[h]        
\begin{center}
\epsfig{file=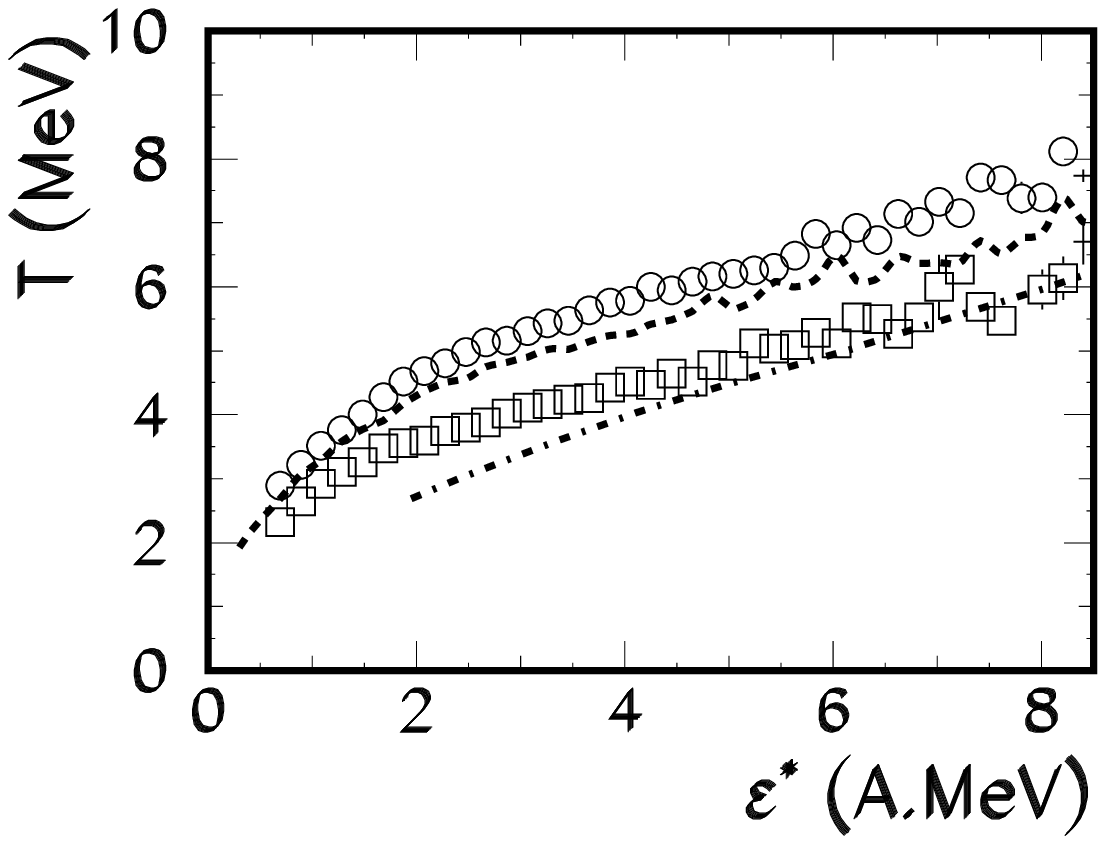,width=0.75\textwidth}
\end{center}
\end{figure}
\noindent
{\bf Fig. 8:} \\
Experimental temperature-excitation energy correlation.
Circles represent the average values calculated with
equations~(\ref{calorimetry}) and (\ref{freeze}) in the {\it SMM-like}
freeze-out hypothesis; squares in the ({\it MMMC-like}) hypothesis.
Dashed and dot-dashed lines represent temperature and excitation energy for
SMM and MMMC filtered predictions respectively, evaluated with 
eq.s~(\ref{calorimetry}) and (\ref{freeze}).
\newpage
\begin{figure}[h]        
\begin{center}
\epsfig{file=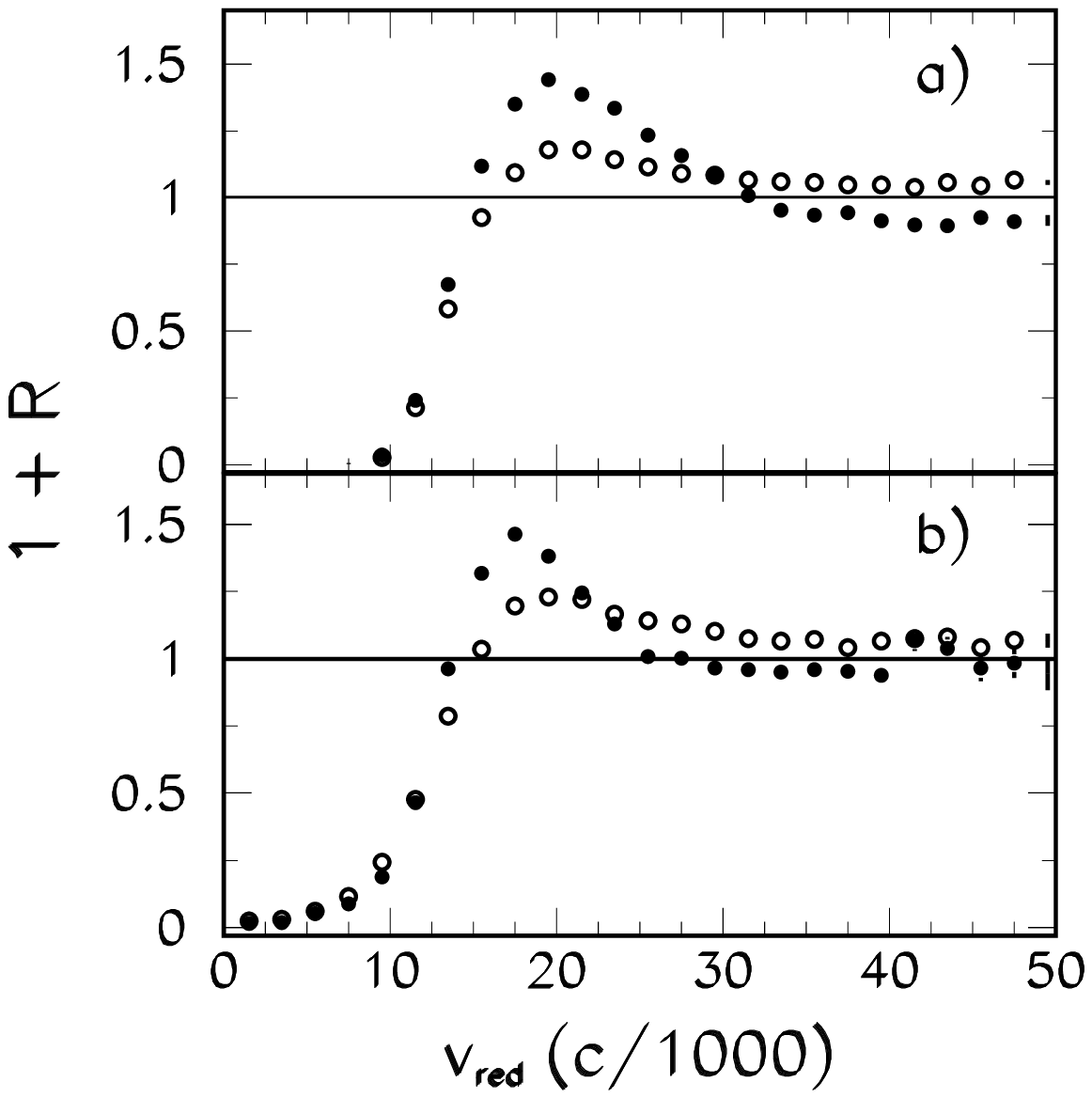,width=0.75\textwidth}
\end{center}
\end{figure}
\noindent
{\bf Fig. 9:} \\
Two-fragment correlation functions of the reduced velocity 
$1 + R(v_{red})$ for 
$4 \le Z \le 15$. Panel~a) shows MMMC calculations and panel~b) SMM 
calculations.
Solid points represent calculations at $\epsilon^* = 4 \ \ A.MeV$, circles at
$6 \ \ A.MeV$.
\newpage
\begin{figure}[h]        
\begin{center}
\epsfig{file=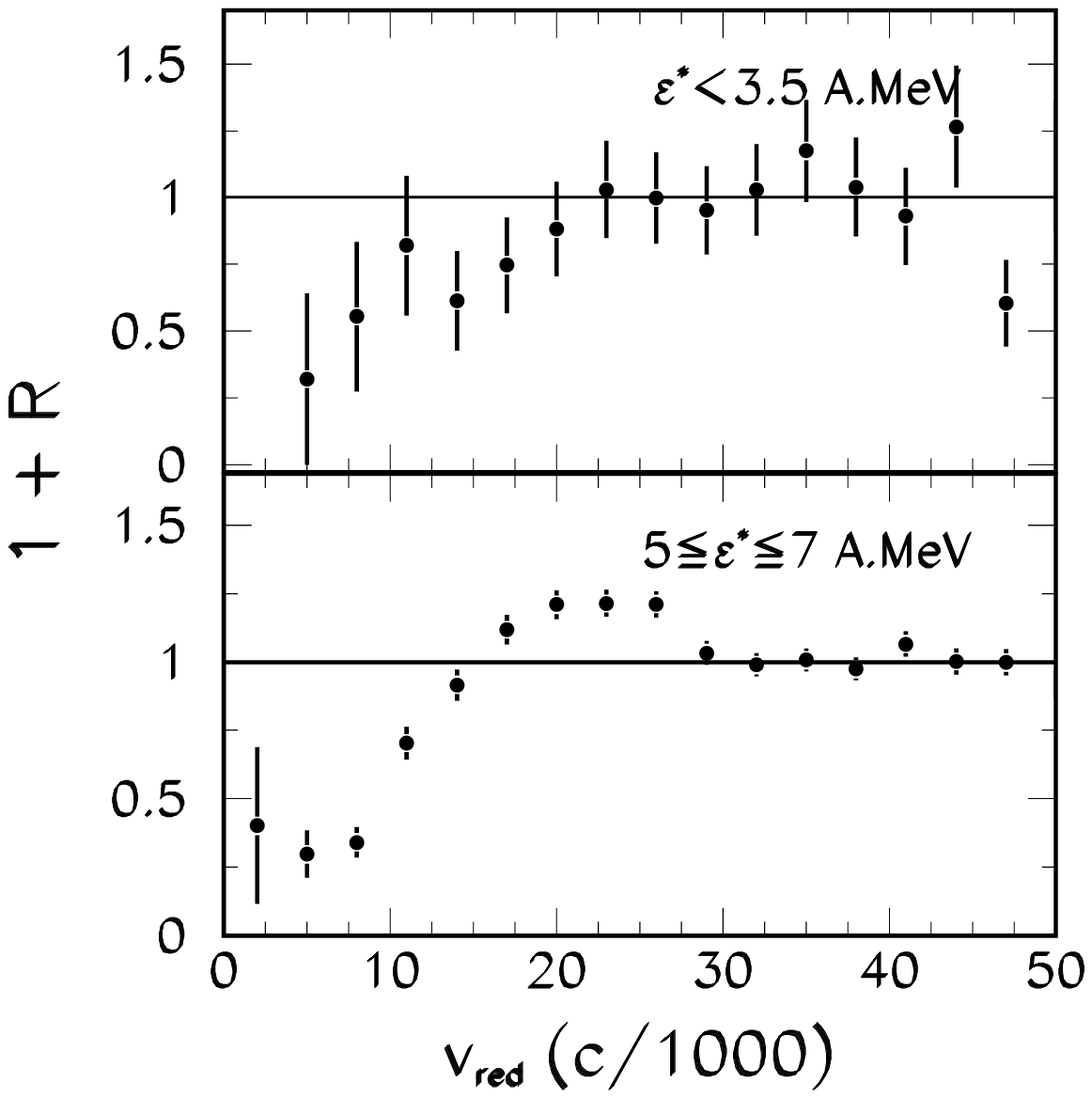,width=0.75\textwidth}
\end{center}
\end{figure}
\noindent
{\bf Fig. 10:} \\
Experimental two-fragment correlation functions of the reduced 
velocity $1 + R(v_{red})$ for $4 \le Z \le 15$ and for
$\epsilon^* < 3.5 \ A.MeV$ 
(upper panel) and for $5 \le \epsilon^* \le 7 A.MeV$ (lower panel).
\newpage
\begin{figure}[h]        
\begin{center}
\epsfig{file=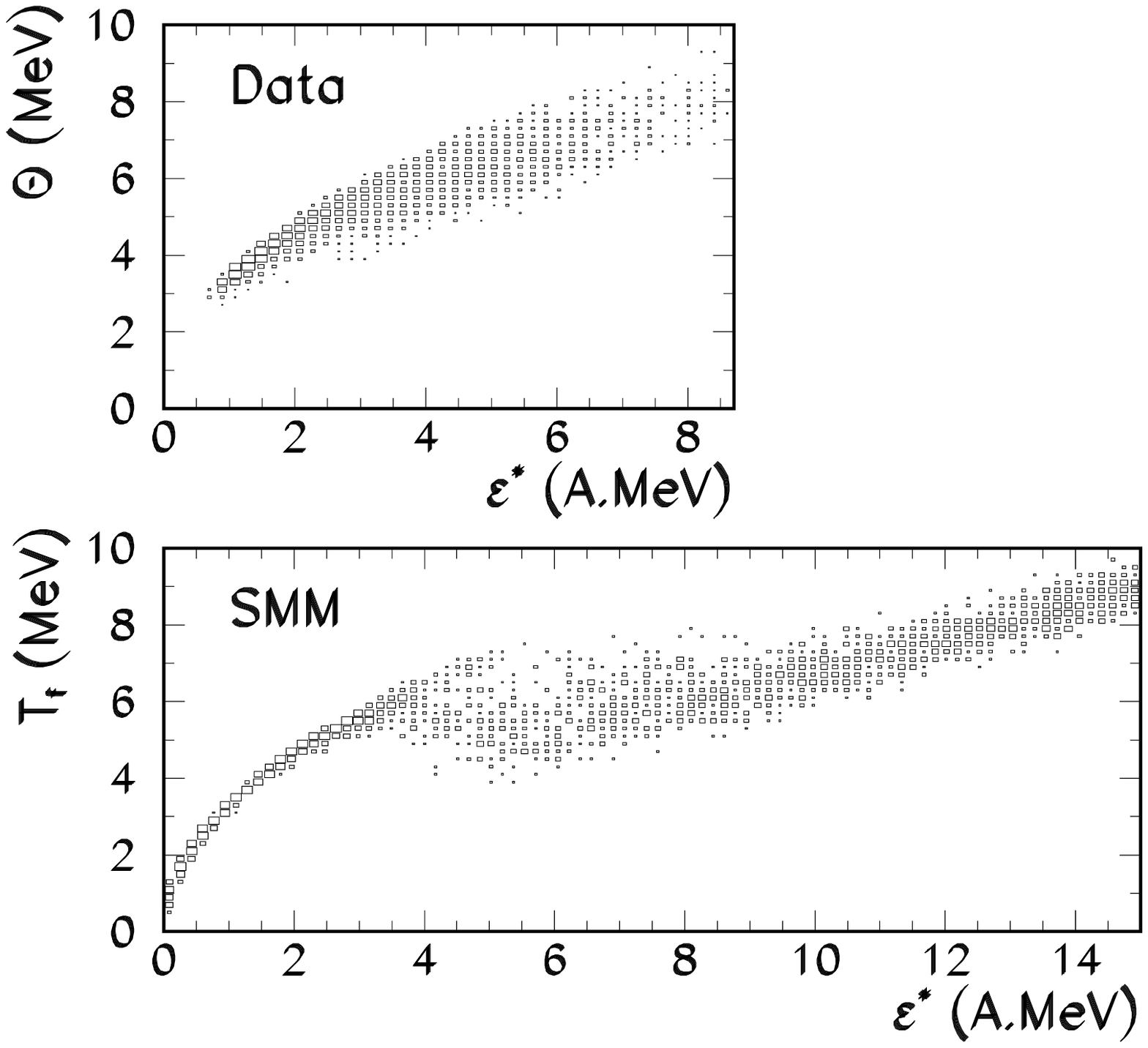,width=0.75\textwidth}
\end{center}
\end{figure}
\noindent
{\bf Fig. 11:} \\
Partition temperature vs. $\epsilon^*$ logarithmic scatter plot.
Upper panel refers to experimental data, lower panel to SMM predictions.
The size of the squares is proportional to the yield.
\newpage
\begin{figure}[h]        
\begin{center}
\epsfig{file=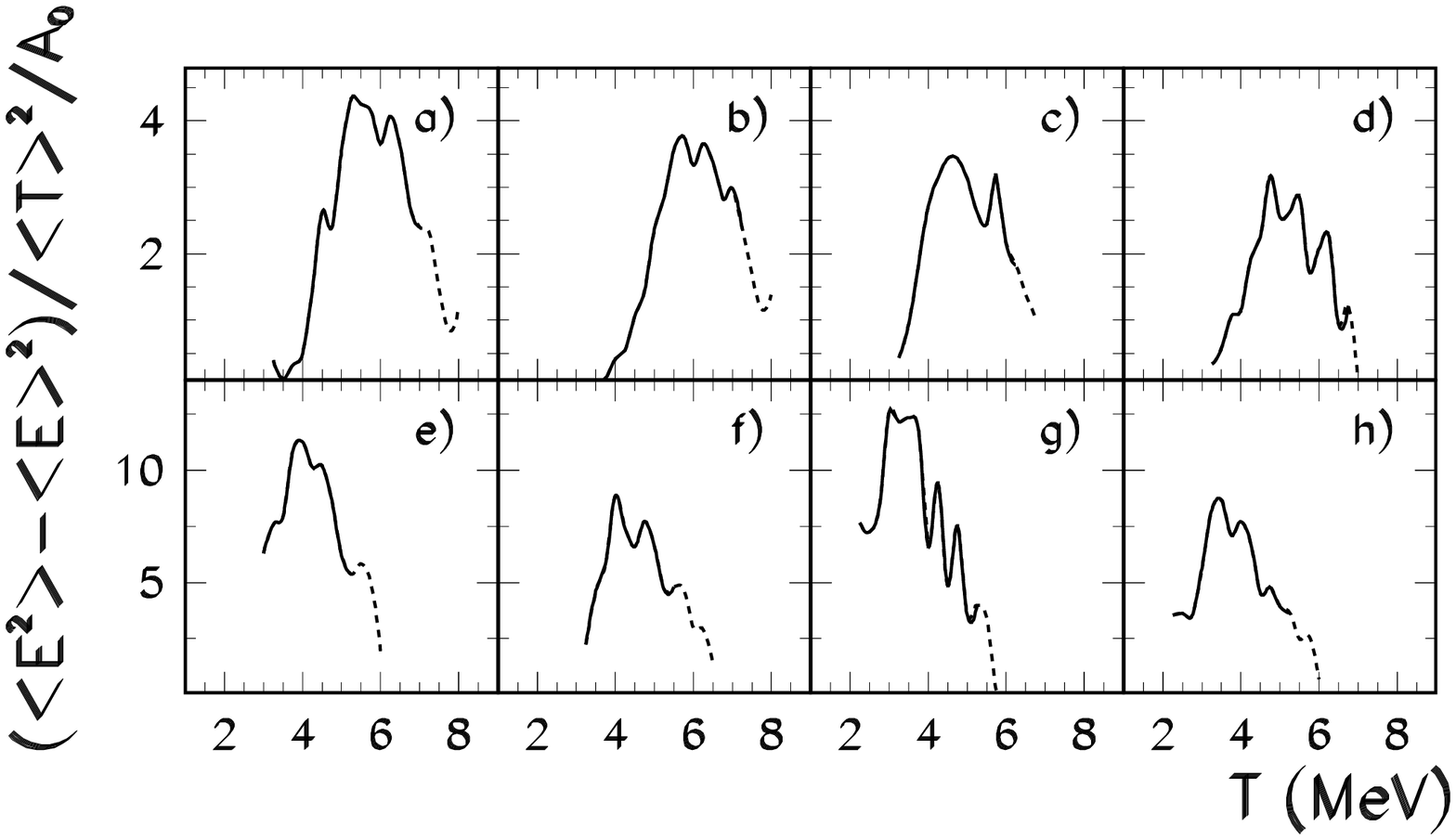,width=0.85\textwidth}
\end{center}
\end{figure}
\noindent
{\bf Fig. 12:} \\
Experimental effective $C_V$ as a function of the 
mean partition temperature.\\
The calculations have been made at a density $\rho_0/3$ in the {\it SMM-like} 
hypothesis (top row) and at a density $\rho_0/6$ in the {\it MMMC-like} 
hypothesis (bottom row).\\
The collective energy has been subtracted from the kinetic energy of the 
detected fragments in panels a), c), e) and g) and it has been included
in panels b), d), f) and h).\\
The level density has been taken $a = f(A)$ in the first and second columns
and $A/8$ in the third and fourth columns.
\newpage
\begin{figure}[h]        
\begin{center}
\epsfig{file=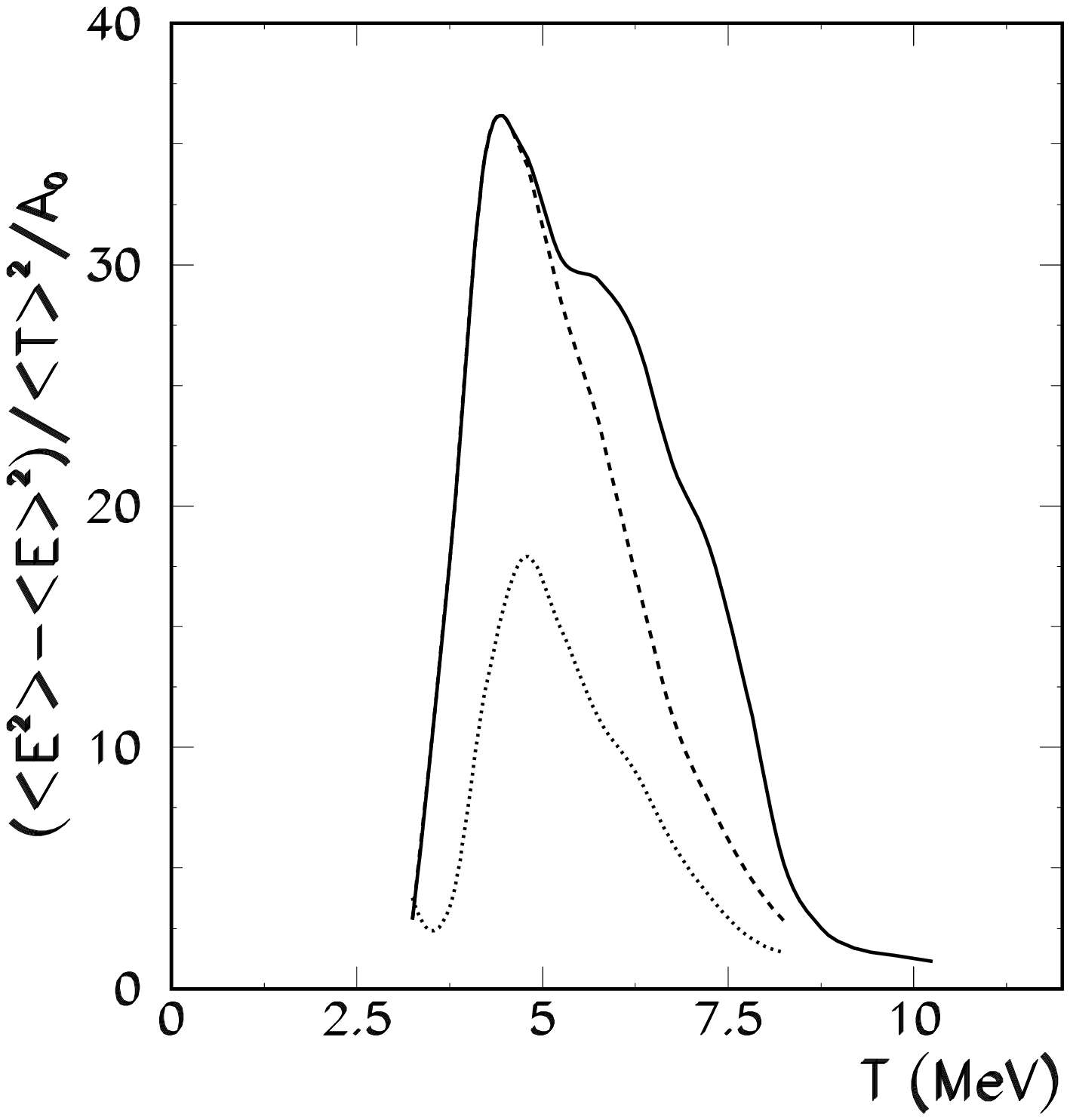,width=0.75\textwidth}
\end{center}
\end{figure}
\noindent
{\bf Fig. 13:} \\
Effective $C_V$ as a function of the mean 
partition temperature for SMM events.\\
The solid line corresponds to $\epsilon^* \le 25 \ A.MeV$ (flat distribution), 
the dashed line to $\epsilon^* < 8 \ A.MeV$ (flat distribution), the dotted 
line to $\epsilon^* < 8 \ A.MeV$ with an exponential probability distribution.
\newpage
\begin{figure}[h]        
\begin{center}
\epsfig{file=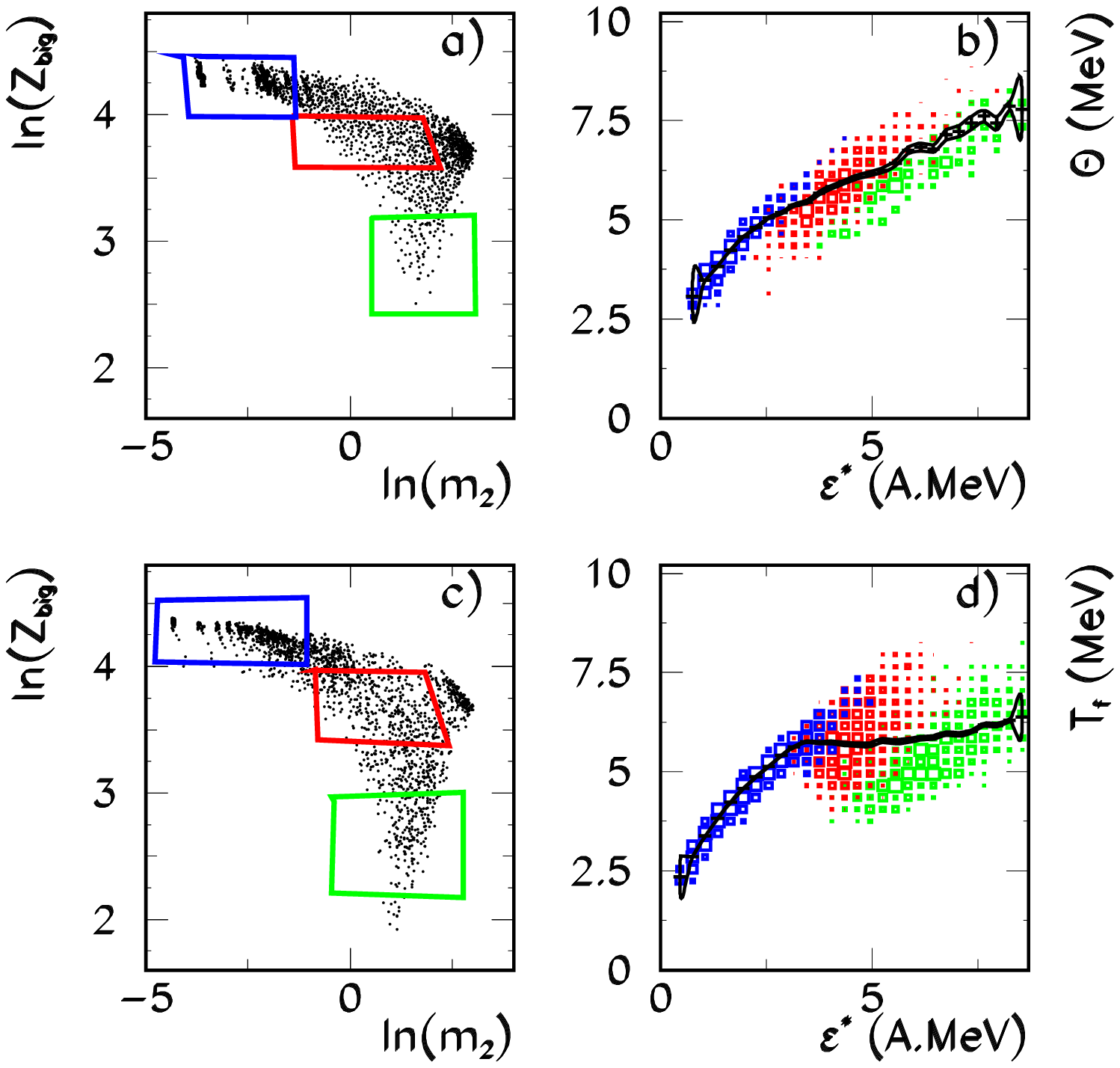,width=0.75\textwidth}
\end{center}
\end{figure}
\noindent
{\bf Fig. 14:} \\
a) Experimental Campi scatter plot. The logarithm of the size
of the largest fragment $ln(Z_{big})$ is plotted versus the logarithm of the
second moment $ln(m_{2})$ (normalized to the charge of the source). Three cuts 
are used to select the upper branch (Cut 1), the lower branch (Cut 3) and the 
central region (Cut 2).\\
b) Experimental $(\Theta, \epsilon^*)$ correlation in Cut 1 (blue), 2 (red)
 and 3 (green). \\
c) SMM Campi scatter plot for filtered events.\\
d) Partition temperature-$\epsilon^*$ correlation in the three cuts for
SMM filtered events.\\
The black line in panels b), d) represents the mean correlation for all the 
events.  The size of the squares is proportional to the yield.
\newpage
\begin{figure}[h]        
\begin{center}
\epsfig{file=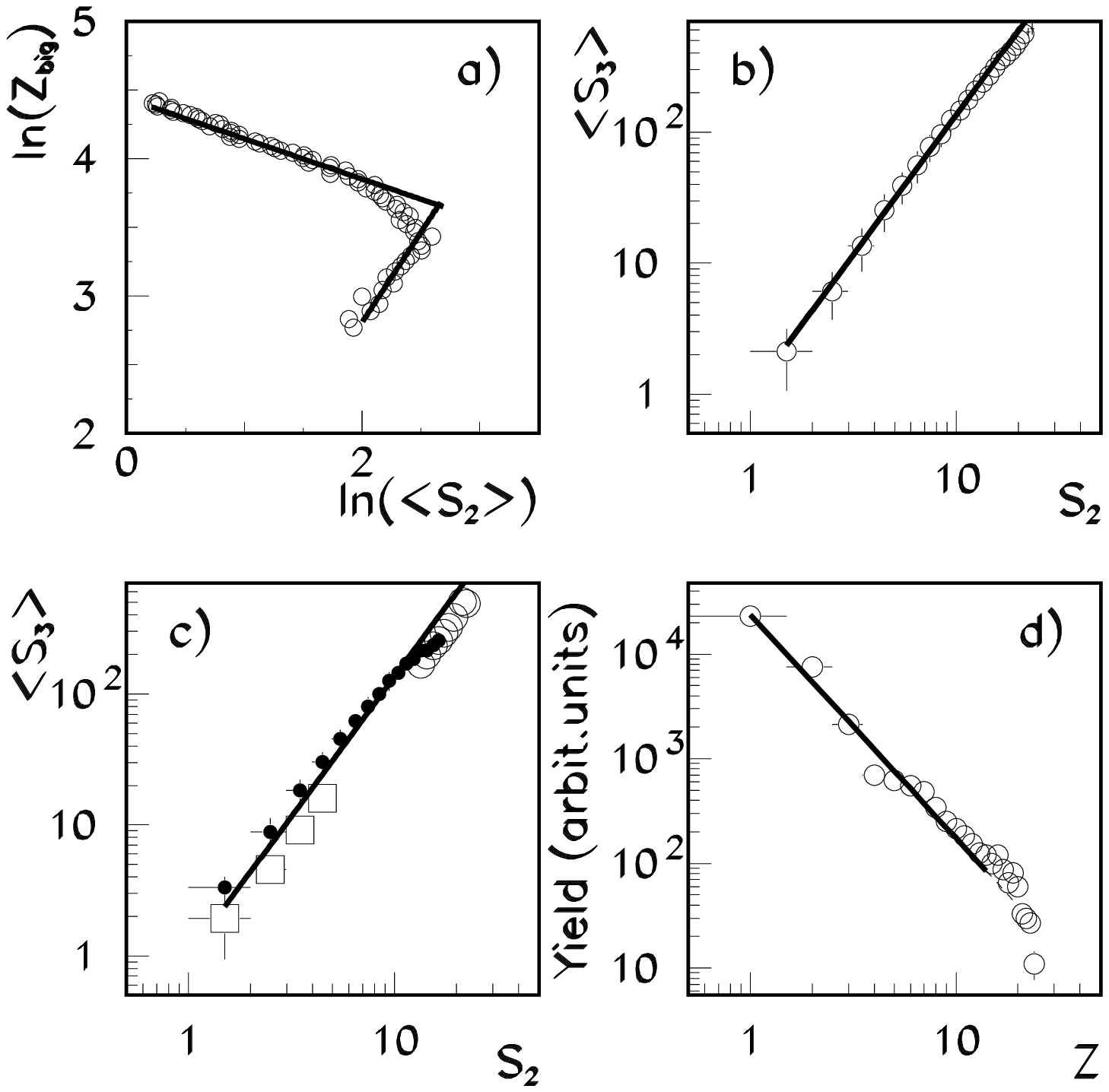,width=0.75\textwidth}
\end{center}
\end{figure}
\noindent
{\bf Fig. 15:} \\
a) $ln(Z_{big})$ versus $ln(S_{2})$. Fission events have been removed.
The lines represent the fit of each branch.\\
b) $S_3$ versus $S_2$. The line is the fit of the correlation.\\
c) $S_3 - S_2$ for the events falling in regions 1 (squares), 2 (points)
and 3 (open circles) of the Campi scatter plot. The line is the fit 
of the whole correlation, shown in panel ~b).\\
d) Charge distribution for events belonging to the Cut 2 of the Campi
scatter plot.
The line represents the power-law fit of the distribution.
\newpage
\begin{figure}[h]        
\begin{center}
\epsfig{file=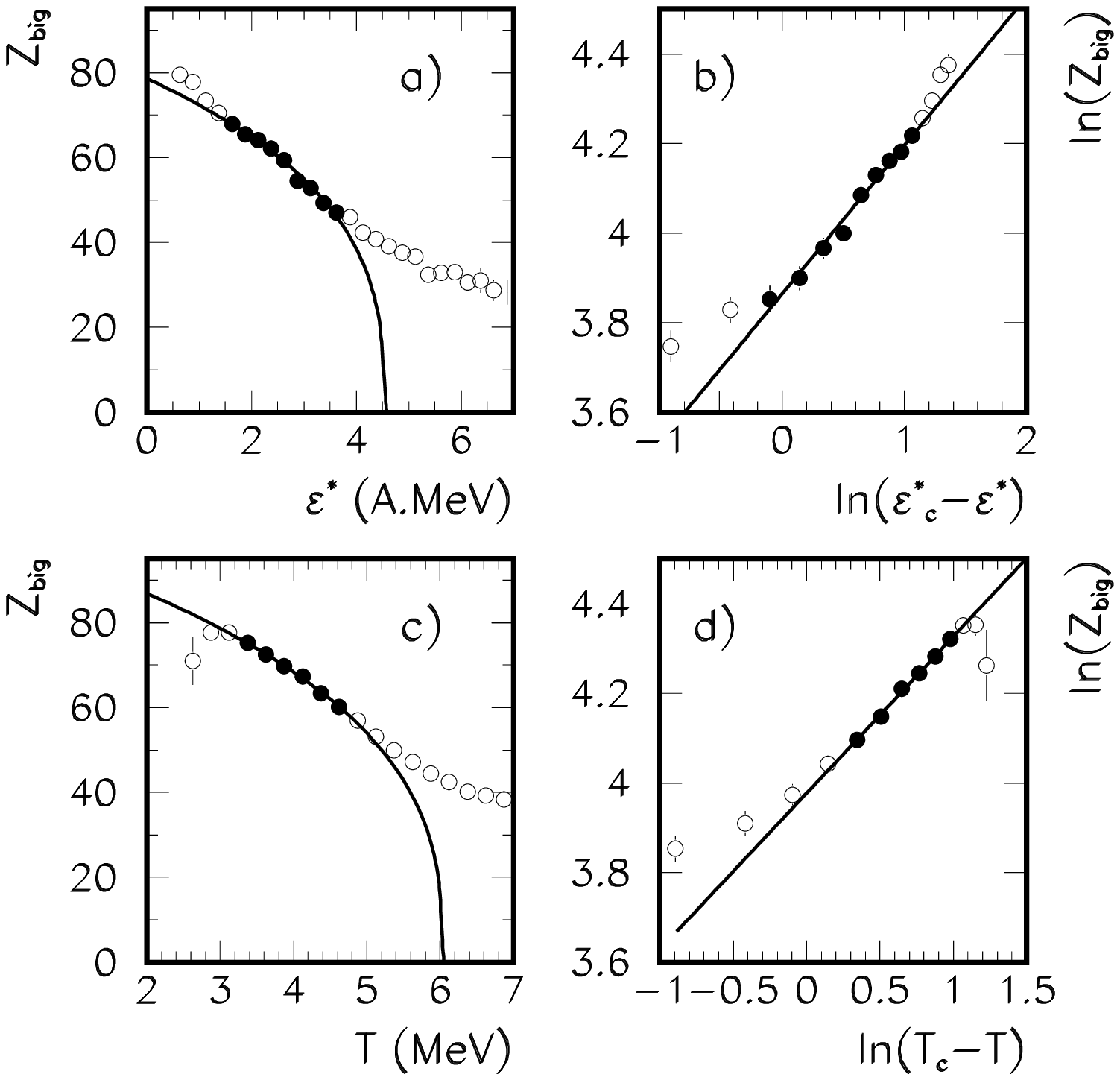,width=0.75\textwidth}
\end{center}
\end{figure}
\noindent
{\bf Fig. 16:} \\
a) $Z_{big}$ vs. $\epsilon^*$ correlation,
b) $ln(Z_{big})$ vs. $ln(\epsilon^*_{crit} - \epsilon^*)$,
c) $Z_{big}$ vs. $T_{crit}$ correlation, d) $ln(Z_{big})$ vs. 
$ln(T_{crit}-T)$. \\
The solid symbols represent the values used for the fits, 
the lines are the fits resulting from equation~(\ref{beta}).
\end{document}